\documentclass[aps,prd,reprint,superscriptaddress,amsmath,amssymb,showkeys]{revtex4-2}
\usepackage[dvipsnames]{xcolor}
\usepackage{dcolumn}
\usepackage{graphicx}
\usepackage{hyperref}
\usepackage{mathrsfs}
\hypersetup{colorlinks=true,linkcolor=blue}

\begin{document}
\title{WIMP Freeze-Out in Diffusive Unimodular Gravity}

\author{Cesar Bonilla}
\email{cesar.bonilla@ucn.cl}
\affiliation{Departamento de Física, Universidad Católica del Norte, Avenida Angamos 0610, Casilla 1280, 1270709 Antofagasta, Chile}

\author{Esteban González}
\email{esteban.gonzalez@ucn.cl}
\affiliation{Departamento de Física, Universidad Católica del Norte, Avenida Angamos 0610, Casilla 1280, 1270709 Antofagasta, Chile}

\author{Carlos Maldonado}
\email{carlos.maldonado@uss.cl}
\affiliation{Facultad de Ciencias, Universidad San Sebasti\'an, Lago Panguipulli 1390, 5501842 Puerto Montt, Chile}

\begin{abstract}
We study a non-standard cosmology (NSC) scenario within the Unimodular Gravity (UG) framework, sourced by a scalar field $\phi$ that undergoes energy diffusion, parametrized by the diffusion parameter $x$, the initial energy densities rate $\kappa\equiv\rho_\phi/\rho_\gamma|_{\text{ini}}$, and the end-of-domination temperature $T_{\text{end}}$. We compare this UG+NSC scenario with standard NSC and $\Lambda$CDM cosmologies for WIMP Dark Matter (DM) production via the freeze-out mechanism. We find that energy diffusion reshapes the allowed $(m_\chi, \langle\sigma v\rangle)$ parameter space, where $m_\chi$ is the DM mass and $\langle \sigma v \rangle$ is the thermally averaged annihilation cross section, opening regions otherwise excluded by DM overproduction in $\Lambda$CDM, and shifting the mass and cross-section ranges accessible to WIMP candidates depending on $x$, $\kappa$, $T_{\text{end}}$, and the barotropic index $\omega$ of $\phi$. As a concrete application, we implement this framework for the Real Singlet Scalar WIMP and test its $(m_\chi, \lambda_{HS})$ parameter space against current direct detection bounds from the LZ experiment, showing that energy diffusion opens previously unconstrained regions in the Higgs-portal coupling $\lambda_{HS}$ and thereby alters the detectability prospects of this benchmark model in future searches.
\end{abstract}


\maketitle

\section{\label{sec:Introduction}Introduction}
The $\Lambda$CDM model is, to date, the most successful cosmological model for the description of our universe. Based on the framework of General Relativity (GR), this model assumes that the universe experiences three different epochs of dominated expansion during its cosmic evolution: radiation at early times, non-relativistic matter in between, and the cosmological constant at late times. Currently, the universe is dominated by the dark sector, composed of cold Dark Matter (cold DM or CDM) and Dark Energy (DE), the latter represented in the model by the cosmological constant $\Lambda$. Together, they represent approximately 30$\%$ and 70$\%$ of the total energy budget of the universe, respectively \cite{Planck:2018vyg}. Nevertheless, even though the $\Lambda$CDM model has been successful in explaining various observational data, such as Type Ia supernovae \cite{Scolnic:2021amr}, measurements of the Hubble parameter \cite{Moresco:2012jh,Zhang:2012mp,Moresco:2015cya}, baryon acoustic oscillations \cite{SDSS:2005xqv}, and the cosmic microwave background \cite{WMAP:2012fli,Planck:2018vyg}, it faces significant observational and fundamental challenges. For the latter, we highlight two of interest for this paper: 1) the nature and evolution of DM and DE remain open questions without definitive answers, and 2) in GR, considering that the vacuum energy density is equivalent to an effective cosmological constant, it is possible to infer a value that differs by 60 to 120 orders of magnitude from the value anticipated by particle physics \cite{Weinberg:1988cp,Carroll:1991mt,Sahni:1999gb,Peebles:2002gy,Padmanabhan:2002ji}, discrepancy known as the cosmological constant problem.

In the task of addressing the nature of DM, we need to consider that, if it is a particle, it must be dark, neutral, non-baryonic, roughly five times more abundant than ordinary matter, non-relativistic, and stable (or sufficiently long-lived). However, no particle within the Standard Model (SM) of particle physics satisfies these requirements simultaneously. Interestingly, theoretical frameworks that were not originally formulated to address the DM problem, such as Supersymmetry (SUSY) \cite{Jungman:1995df} and string theory \cite{King:2006cu}, naturally contain a particle that can play the role of a DM candidate, capable of accounting for the cosmological and gravitational observational evidence of DM. More generally, a wide variety of extensions of the SM have been shown to accommodate such candidates, ranging from axions to sterile neutrinos and other extended sectors (see, e.g., \cite{Arbey:2021gdg,Cirelli:2024ssz,Bozorgnia:2024pwk} for comprehensive reviews). Rather than constituting an exhaustive or uniquely motivated list, these candidates are commonly organized according to their production mechanism and the strength of their coupling to the SM, with Weakly Interacting Massive Particles (WIMPs) \cite{Steigman:1984ac, Bertone:2004pz, Arcadi:2017kky, Roszkowski:2017nbc, Arcadi:2024ukq, Singh:2024wdn} and Feebly Interacting Massive Particles (FIMPs) \cite{Bernal:2017kxu, Chu:2011be, Hall:2009bx} representing two of the most extensively studied paradigms. We focus on the former: besides emerging naturally in well-motivated SM extensions such as SUSY, WIMPs are predicted to have electroweak scale interactions that yield a thermal relic abundance in agreement with observations (the so-called ``WIMP miracle''), and, crucially, this same interaction strength makes them within the reach of a robust experimental program, including direct detection \cite{XENON:2018voc,LZ:2022lsv}, indirect detection\cite{Gaskins:2016cha,Bringmann:2012ez}, and collider searches~\cite{Kahlhoefer:2017dnp}. It is precisely this rich detectability landscape that motivates our focus on WIMP DM in what follows.

WIMPs are massive particles, possibly in the GeV--TeV range, characterized by interacting only through the electroweak force and gravity, and are typically produced via the freeze-out mechanism. This mechanism implies that a non-negligible initial DM particle population is kept in thermal equilibrium with the SM bath. Nevertheless, as the universe expands and cools, this equilibrium is eventually broken, leaving behind the currently observed DM relic density, $\Omega_{c}h^{2}=0.12$ according to Planck \cite{Planck:2018vyg}, where $\Omega_{c}$ is the DM density parameter and $h$ is the reduced Hubble parameter.

To be consistent with observations within the $\Lambda$CDM framework, the total thermally averaged DM annihilation cross section must be approximately $\langle\sigma v\rangle_0 = \text{few} \times 10^{-9}\;\text{GeV}^{-2}$ \cite{Steigman:2012nb}. However, no conclusive results have been obtained to date in the search for WIMP DM. This tension is particularly relevant given that the DM relic density depends heavily on the cosmological era in which it is established. In the standard $\Lambda$CDM picture, this relic density is assumed to be set during the radiation-dominated era. Hence, if the expansion history of the universe differs from this standard assumption, the resulting DM abundance can be significantly modified. Along this line, several interesting cosmological scenarios include an additional field ($\phi$) in the early universe. This field can generate different domination eras and decay into SM particles, leading to entropy injection into the SM bath, which enlarges the available parameter space for DM and modifies its production if it occurs during these non-standard eras \cite{Giudice:2000ex,Salati:2002md,Pallis:2004yy,Gelmini:2006pw,Gelmini:2006pq,Randall:2015xza, Tenkanen:2016jic, Hamdan:2017psw, DEramo:2017ecx, DEramo:2017gpl, Visinelli:2017qga, Drees:2018dsj, Bernal:2018ins, Bernal:2018kcw, Poulin:2019omz, Maldonado:2019qmp, Arias:2019uol, Bernal:2019mhf, Cosme:2020mck, Arcadi:2021doo,Arias:2021rer, Bernal:2022wck, Haque:2023yra, Silva-Malpartida:2023yks,Barman:2024mqo,Ghosh:2023tyz, Silva-Malpartida:2024emu, Arcadi:2024tib}. These alternative scenarios, known as Non-Standard Cosmologies (NSCs), offer new avenues for DM detection and may reopen regions of the parameter space previously excluded in the $\Lambda$CDM model, potentially revealing viable DM candidates and production mechanisms. Therefore, if DM is experimentally detected, its particle physics properties—such as mass, interactions with SM particles, and coupling constants—must be determined in a way that is consistent with the underlying cosmological framework. Conversely, it is also necessary to explore NSC scenarios in which these properties can be consistently accommodated.

Focusing on the second challenge, Unimodular Gravity (UG) arises as an alternative theory to GR in which the cosmological constant is merely an integration constant. This theory was proposed by Einstein himself~\cite{Einstein:1919gv,lorentz1952principle}, by writing the trace-free part of his GR field equations with a fixed determinant of the metric ($\sqrt{-g}=1$), which simplifies some calculations in certain circumstances~\cite{Bufalo:2015wda}. Nowadays, UG can be described in different equivalent ways~\cite{Anderson:1971pn,vanderBij:1981ym,Buchmuller:1988wx,Unruh:1988in,Henneaux:1989zc,Ng:1990xz,Finkelstein:2000pg,Ellis:2010uc}, with the most usual one being based on invariance under a restricted group of diffeomorphisms, in which the determinant of the metric can be set equal to a nondynamical 4-form according to $\sqrt{-g}=\epsilon_{0}$. This condition breaks the diffeomorphism invariance down to volume preserving diffeomorphisms~\cite{Bufalo:2015wda}, i.e., those generated by vector fields $\xi^{\mu}$, satisfying $\nabla_{\mu}\xi^{\mu}=0$. The Noether theorem associated with this symmetry implies a modified conservation law for matter fields. This is because diffeomorphism invariance of the matter action generated by an arbitrary vector field implies the conservation law for the energy-momentum tensor $\nabla^{\mu}T_{\mu\nu}=0$, but, in UG, the vector fields that generate the symmetries are no longer arbitrary. One of the advantages of the UG formalism is that, on shell, Bianchi identities give rise to the same Einstein field equations of GR but with the cosmological constant arising as an integration constant~\cite{Corral:2020lxt}. Even more, the vacuum energy density of quantum fields can be removed from the field equations by rescaling an additional component of the energy-momentum tensor that appears from the restricted invariance, resulting in the UG formalism not suffering the cosmological constant problem~\cite{Smolin:2009ti}.

It is important to mention that the modified version of the conservation law of matter obtained in UG implies, in other words, a nonconservation of the energy-momentum tensor, contrary to what happens in GR. While this can be seen as a disadvantage of the UG gravity formalism, in fact, it provides a suitable setup to reconcile gravitation with the energy-momentum nonconservation that arises from quantum collapse or quantum gravity discreteness~\cite{Josset:2016vrq,Perez:2017krv,Perez:2018wlo}. Also, this nonconservation leads to an effective variable cosmological constant that can drive the recent acceleration in the universe expansion~\cite{Josset:2016vrq}, can provide a resolution to the Hubble tension~\cite{Perez:2020cwa}, and might explain the low spin of black holes detected via gravitational waves \cite{Perez:2019gyd}. Finally, the nonconservation of the energy-momentum tensor opens new windows for exploring physics beyond the $\Lambda$CDM model. For example, in Ref.~\cite{Corral:2020lxt}, a cosmological UG model has been explored that successfully describes the joint analysis of SNe Ia and observational Hubble parameter data (OHD) in comparison with the $\Lambda$CDM model, providing insights that a very small but nontrivial energy nonconservation is compatible with the model.

In Refs.~\cite{Gonzalez:2024dtb} and \cite{Gonzalez:2024rhs}, a novel NSC scenario was studied in which the new field $\phi$ experiences dissipative processes in the form of a bulk viscosity within the framework of Eckart's theory. This NSC was applied to WIMP and FIMP DM candidates. The most important finding is that the parameter space shows significant deviations from the classical NSC scenario, yielding new regions to search for these DM candidates. In particular, for certain combinations of the free parameters, the authors found large regions where the model can successfully accommodate DM and reproduce the currently observed relic density. Motivated by this, this paper aims to propose a novel NSC scenario within the framework of UG, where this new field $\phi$ instead experiences energy diffusion. While bulk viscosity in the dissipative case affects the acceleration and conservation equations, the diffusion function in UG also impacts the first Friedmann equation; therefore, we expect significant deviations from the standard NSC scenario. This novel NSC is studied for WIMPs DM candidates, with a particular application to the Real Singlet Scalar DM, to investigate how energy diffusion in the scalar field leaves imprints on WIMP DM production and its relic density.

This paper is organized as follows: In Section~\ref{sec:ClassicNSC}, we briefly describe the classical NSC scenario, and in Subsection~\ref{subsec:WIMPS}, we apply this classical scenario to WIMP DM candidates. In Section~\ref{sec:Unimodular}, we outline the UG formalism and present the equations that govern the evolution of a universe described by the spatially flat Friedmann-Lemaître-Robertson-Walker (FLRW) metric. In Section~\ref{sec:UnimodularNSC}, we introduce the novel NSC scenario within the framework of UG with energy diffusion, comparing it with the classical NSC in Subsection~\ref{subsec:Comparison}. In Subsections~\ref{subsec:WIMPParameters} and~\ref{subsec:RSSParameters}, this novel NSC scenario is studied for WIMP DM candidates in a general way and for the particular case of Real Singlet Scalar WIMP DM, respectively. Finally, in Section~\ref{sec:Conclusions}, we present our conclusions and final discussions.

\section{\label{sec:ClassicNSC}Non-standard cosmologies}
In the $\Lambda$CDM model, at early times the Universe is dominated by radiation, during which the DM abundance evolves until it freezes out and reaches its present-day value. Alternative cosmological histories can be obtained by introducing a new field $\phi$, which dominates these early times and decays into the SM plasma \cite{Giudice:2000ex, Hamdan:2017psw, DEramo:2017ecx, DEramo:2017gpl, Visinelli:2017qga, Drees:2018dsj, Bernal:2017kxu, Bernal:2018ins, Maldonado:2019qmp, Arias:2019uol, Bernal:2019mhf, Arias:2021rer, Bernal:2022wck, Silva-Malpartida:2024emu}. The evolution of the energy density of this new component, $\rho_\phi$, and the entropy density of the SM, $s$, is determined from the conservation equations, yielding
\begin{equation}
    \dot{\rho_\phi}+3H\rho_\phi(\omega+1)=-\Gamma_\phi \rho_\phi,
\end{equation}
\begin{equation}
    \dot{s}+3Hs=\frac{\Gamma_\phi \rho_\phi}{T},
    \label{eqentropy}
\end{equation}
where the ``dot'' denotes the derivative with respect to cosmic time $t$, $H$ is the Hubble parameter, $\Gamma_\phi$ is the decay rate of the field, and $\omega$ is the barotropic index of this new state $\phi$. The entropy of the SM is defined by the temperature $T$ of the photons as
\begin{equation}
    s=\frac{\rho_\gamma+p_\gamma}{T}=\frac{2\pi^2}{45}g_{\star s}(T)T^3,
    \label{entropy}
\end{equation}
with $\rho_\gamma$ and $p_\gamma$ being the radiation energy density and its pressure, and $g_{\star s}(T)$ the effective number of relativistic degrees of freedom that contribute to the entropy density. In this sense, from the temperature of the photons, it is possible to write the radiation energy density as
\begin{equation}
    \rho_\gamma=\frac{\pi^2}{30}g_\star(T)T^4,
\end{equation}
where $g_\star(T)$ is the effective number of relativistic degrees of freedom that contribute to the radiation energy density. Finally, the temperature of the SM can be tracked by replacing Eq.~\eqref{entropy} into Eq.~\eqref{eqentropy}, being obtained
\begin{equation}
    \frac{dT}{da}=\left(1+\frac{T}{3g_{\star s}(T)}\frac{dg_{\star s}(T)}{dT}\right)^{-1}\left[-\frac{T}{a}+\frac{\Gamma_\phi \rho_\phi }{3Hsa}\right],
    \label{eqtemp}
\end{equation}
where $a$ is the scale factor. The Hubble parameter $H\equiv\dot{a}/a$ is given, for a flat FLRW universe, by the Friedmann equation
\begin{equation}\label{Hubbleeq}
    H^2=\frac{\sum_i \rho_i}{3M_p^2},
\end{equation}
where $M_p$ is the reduced Planck mass, and the sum runs over all the fluids that compose the universe. It is important to mention that, to be consistent with observations of the Universe after Big Bang Nucleosynthesis (BBN), this new state $\phi$ must decay at a temperature $T_\text{end} \gtrsim 4$ MeV. This temperature is determined by the total decay width $\Gamma_\phi$, according to the expression
\begin{equation}
    T_\text{end}^4= \frac{90}{\pi^2 g_\star(T_\text{end})}M_p^2\Gamma_\phi^2.
\end{equation}

\begin{figure}
\centering
\includegraphics[width=0.48\textwidth]{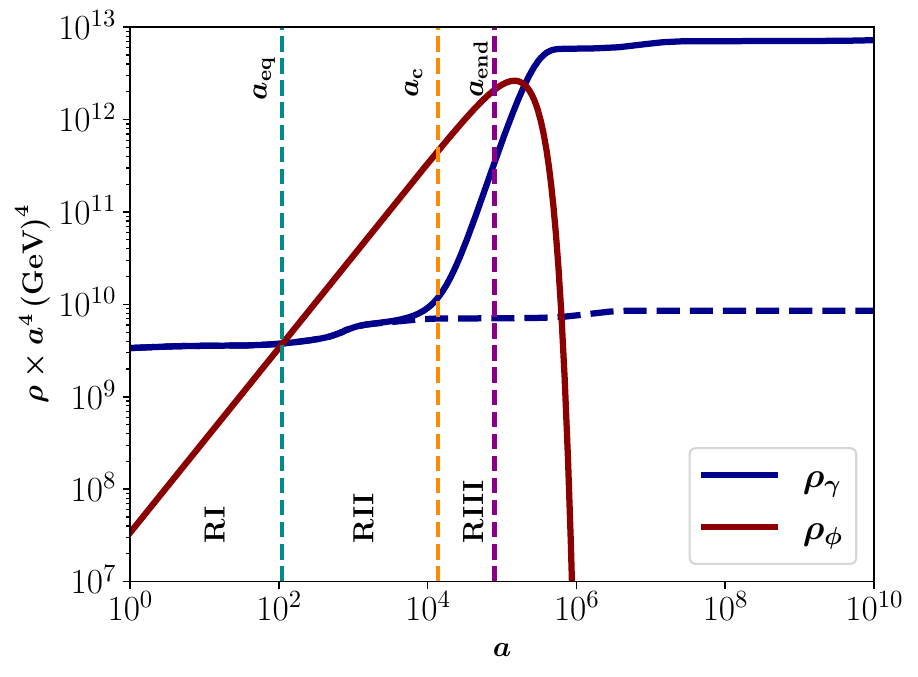}
\caption{Comparison between $\Lambda$CDM and NSC scenarios for $\omega=0$, $T_\text{end}=7\times10^{-3}\;\text{GeV}$, $\kappa=10^{-2}$, and $T_i=100$ GeV. The blue and red solid lines correspond to the energy density of radiation and $\phi$ in the NSC scenario, respectively. The dashed blue line corresponds to the energy density of radiation in $\Lambda$CDM. The cyan, orange, and magenta dashed lines correspond to $a_\text{eq}$, $a_\text{c}$, and $a_\text{end}$, respectively.}
\label{Comparacionstdnsc}
\end{figure}

Figure \ref{Comparacionstdnsc} shows the evolution of the energy densities in the NSC scenario compared to $\Lambda$CDM as a function of the scale factor for $\omega=0$, $T_\text{end}=7\times10^{-3}$ GeV, $\kappa=10^{-2}$, and $T_i=100$ GeV, with $\kappa\equiv\rho_{\phi_0}/\rho_{\gamma_0}$, i.e., the initial ratio between energy densities of $\phi$ and radiation. Three regions of interest can be identified. Region I (RI), defined by $a < a_\text{eq}$ (cyan dashed line), is characterized by radiation domination over $\phi$ and is possible only with $\kappa<1$. Region II (RII), defined by $a_\text{eq} < a < a_\text{c}$ (orange dashed line), corresponds to a $\phi$ dominated Universe. Region III (RIII), defined by $a_\text{c} < a < a_\text{end}$, is also dominated by $\phi$, with its decays into the SM plasma becoming effective until the temperature $T_\text{end}$ is reached. After the decay of $\phi$, i.e., for $a > a_\text{end}$, the $\phi$ field has fully decayed, and the $\Lambda$CDM scenario is restored.

The evolution of the DM number density, $n_\chi$, is described by the following Boltzmann equation
\begin{equation}
\dot{n_\chi}+3Hn_\chi=-\langle\sigma v\rangle\left(n_\chi^2-n_{\chi_{eq}}^2\right),
\label{dmboltzmann}
\end{equation}
where $\langle \sigma v \rangle$ is the thermally averaged cross section for DM annihilation/production, and $n_{\chi}^{\text{eq}}$ is the equilibrium number density, defined as $n_{\chi}^{\text{eq}} = m_\chi^2 T K_2(m_\chi/T)/\pi^2$, with $K_2$ denoting the modified Bessel function of the second kind and $m_\chi$ the DM mass. The energy density of DM is given in terms of its number density as $\rho_\chi = m_\chi n_\chi$, which contributes to the Hubble parameter in Eq.~\eqref{Hubbleeq}.

As mentioned above, NSC scenarios introduce an early $\phi$-dominated era, leading to an enhancement of the radiation energy density, i.e., an entropy injection, which leaves imprints on DM production.

\subsection{\label{subsec:WIMPS}WIMPs in non-standard cosmologies}
The class of particles known as WIMPs is motivated by the ``WIMP miracle'', which states that, for a DM mass of order 100 GeV, the thermally averaged annihilation cross section must be of the order of the electroweak scale in order to reproduce the observed DM relic density. Moreover, for a wide range of DM masses, the required thermally averaged annihilation cross section in the $\Lambda$CDM scenario is $\langle \sigma v \rangle_0 = \text{few} \times 10^{-9}\;\text{GeV}^{-2}$. Also, in this scenario, $\langle \sigma v \rangle > \langle \sigma v \rangle_0$ leads to an underabundance of DM, since stronger interactions keep the particles in thermal equilibrium for a longer time and delay freeze-out. This case is compatible with multi-component DM. On the other hand, $\langle \sigma v \rangle < \langle \sigma v \rangle_0$ results in an overabundance of DM particles due to weaker interactions, which lead to an earlier freeze-out.

To track the evolution of DM, it is convenient to define the yield as $Y \equiv n_\chi/s$ and the variable $x \equiv m_\chi/T$. An analytical solution to Eq. \eqref{dmboltzmann} is obtained in the limit $Y\gg Y_\text{eq}$, yielding
\begin{equation}
    Y\propto \frac{1}{m_\chi J(x_\text{fo})},
\end{equation}
with $J = \int_{x_\text{fo}}^\infty x^{-2}\langle \sigma v \rangle(x),dx$, an integral over $x$, where the thermally averaged cross section depends on this variable. In this case, $x_\text{fo}$ represents the time at which the DM candidate freezes out, since its interactions can no longer compete with the expansion of the Universe. 

\begin{figure}
\centering
\includegraphics[width=0.48\textwidth]{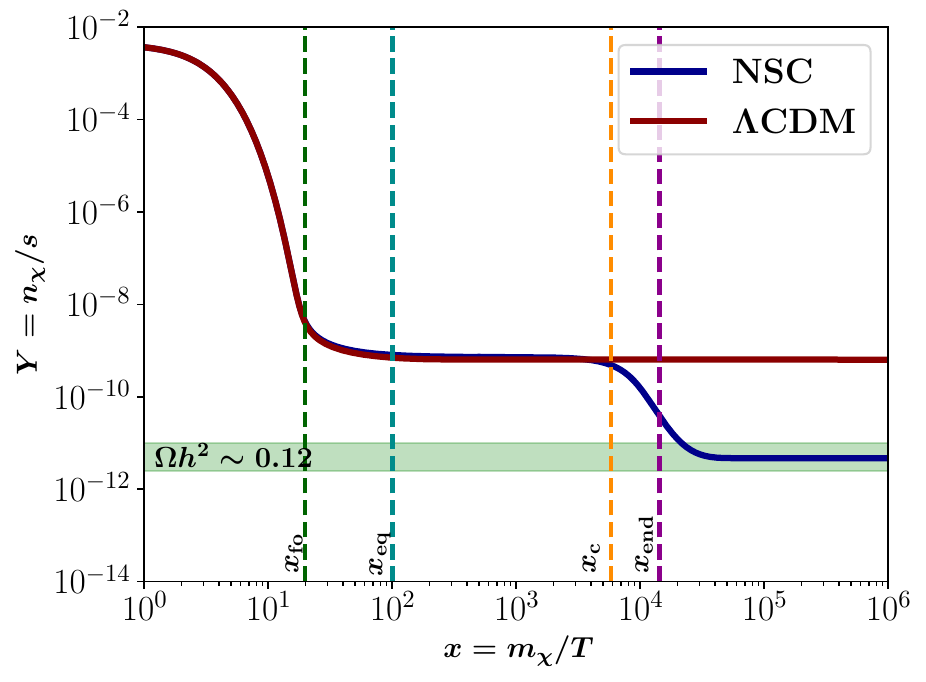}
\caption{Comparison between $\Lambda$CDM and NSC scenarios for the Yield of WIMP DM candidate for $\kappa=10^{-2}$, $\omega=0$, $T_\text{end}=7\times10^{-3}$ GeV, $m_\chi=100$ GeV, and $\langle\sigma v\rangle=10^{-11}$ GeV$^{-2}$. The blue and red solid lines correspond to the Yield of DM in the NSC scenario and $\Lambda$CDM, respectively. The green, cyan, orange, and magenta dashed lines correspond to $x_\text{fo}$, $x_\text{eq}$, $x_\text{c}$, and $x_\text{end}$, respectively. The green strip corresponds to the current DM relic density.}
\label{Comparacionstdnscdm}
\end{figure}

The inclusion of $\phi$ modifies the evolution of the early Universe, thereby changing the time at which DM is established. The entropy injection into the SM plasma dilutes the yield and, consequently, the DM relic density. As a result, lower values of $\langle \sigma v \rangle$, which are excluded in the $\Lambda$CDM scenario due to the overproduction of DM, become allowed in the NSC scenario. This is illustrated in Figure \ref{Comparacionstdnscdm} for $\langle \sigma v \rangle = 10^{-11}\,\text{GeV}^{-2}$ and $m_\chi=100$ GeV, where the red and blue lines correspond to the $\Lambda$CDM and NSC scenarios, respectively. The NSC parameters in this case are $\omega = 0$, $\kappa = 10^{-2}$, and $T_\text{end} = 7 \times 10^{3}\;\text{GeV}$. These DM parameters are excluded in $\Lambda$CDM due to the overproduction of DM; however, in the NSC scenario, they are allowed due to entropy injection into the SM plasma from $\phi$ decays. Note that the WIMP DM candidate undergoes freeze-out in RI of the NSC scenario (i.e., before $a_\text{eq}$), so the DM abundance is set as in $\Lambda$CDM. However, when $\phi$ decays inject entropy into the SM bath (between the orange and magenta dashed lines), the yield is diluted until it matches the observed DM relic density for these parameters.

\section{\label{sec:Unimodular}Unimodular Gravity}
The modern definition of UG is based on a restricted group of diffeomorphisms, in which the determinant of the metric can be set equal to a fixed scalar density $\epsilon_{0}$, i.e., $\sqrt{-g}=\epsilon_{0}$~\cite{Bufalo:2015wda}. This condition can be introduced in the Einstein-Hilbert action through a Lagrange multiplier $\lambda$, which provides a fixed-volume element in spacetime, on shell, as follows
~\cite{Padilla:2014yea,Corral:2020lxt}:
\begin{align}\label{UGaction}\notag
S\left[g_{\mu\nu},\lambda,\Psi\right] &= \frac{1}{2\alpha}\int{d^4 x\sqrt{-g}\left[R-2\lambda\left(1-\frac{\varepsilon_0}{\sqrt{-g}}\right)\right]}\\&\quad+\int{d^4 x\sqrt{-g}\,\mathscr{L}_m\left[g_{\mu\nu},\Psi\right]},
\end{align}
where $\alpha=8\pi G/c^{4}$, with $G$ the gravitational constant and $c$ the speed of light, $R$ is the Ricci scalar, $g_{\mu\nu}$ is
the metric tensor of the four-dimensional spacetime, $\mathscr{L}_m$ is the matter Lagrangian, and $\Psi$ represents the matter fields. By performing arbitrary variations on the above action with respect to $g_{\mu\nu}$, $\lambda$, and $\Psi$, the following set of equations is obtained:
\begin{eqnarray}
R_{\mu\nu}-\frac{1}{2}R g_{\mu\nu}+\lambda g_{\mu\nu}&=&\alpha T_{\mu\nu}, \label{originalUGEFE} \\
\sqrt{-g}&=&\epsilon_{0}, \\ \label{UGcondition}
\dfrac{\delta\mathscr{L}_{m}}{\delta\Psi}&=&0,
\end{eqnarray}
where $R_{\mu\nu}$ and $T_{\mu\nu}$ are the Ricci tensor and energy-momentum tensor, respectively. Note that Eq.~\eqref{originalUGEFE} is very similar to Einstein's field equations
\begin{equation}\label{EFE}
R_{\mu\nu}-\frac{1}{2}R g_{\mu\nu}+\Lambda g_{\mu\nu}=\alpha T_{\mu\nu},
\end{equation}
with the difference that $\lambda$ in UG can be interpreted as a cosmological function instead of a cosmological constant $\Lambda$ like in GR. 

To see the difference between GR and UG more clearly, we can take the trace of Eq.~\eqref{originalUGEFE}, which allows us to obtain $\lambda$, taking the form
\begin{equation}\label{sollambda}
\lambda=\frac{R+\alpha T}{4},
\end{equation}
where $T$ is the trace of the energy-momentum tensor. The above equation can be substituted into Eq.~\eqref{originalUGEFE}, yielding
\begin{equation}\label{TFEE}
R_{\mu\nu}-\frac{1}{4}Rg_{\mu\nu}=\alpha\left(T_{\mu\nu}-\frac{1}{4}Tg_{\mu\nu}\right),
\end{equation}
an equation that corresponds to the trace-free part of Einstein's field equations \eqref{EFE}~\cite{Ellis:2013uxa}. If we take the covariant divergence of Eq.~\eqref{TFEE}, and use the Bianchi identity, it is obtained
\begin{equation}\label{nablaTFEE}
\frac{1}{4}\nabla_{\nu}\left(R+\alpha T\right)=\alpha \nabla^{\mu}T_{\mu\nu}.
\end{equation}
Then, by substituting Eq.~\eqref{sollambda} into Eq.~\eqref{nablaTFEE}, we obtain a first integral of motion given by
\begin{equation}\label{lambda}
\lambda=\Lambda+\alpha\int_{l}{J},
\end{equation}
where $\Lambda$ is an integration constant and $J= \nabla^{\mu}T_{\mu\nu}$ represents the energy–momentum current violation, which is integrated over some arbitrary path $l$~\cite{LinaresCedeno:2020uxx}. Note that Eq.~\eqref{lambda} makes clear that $\lambda$ is a function instead of a cosmological constant. In fact, in UG, $\Lambda$ arises as an integration constant which coincides with the cosmological constant of GR when $\nabla^{\mu}T_{\mu\nu}=0$, i.e., when the energy-momentum tensor is conserved. Therefore, we can define $Q(x)=\int_{l}{J}$, which is an arbitrary function of some variable $x$ that measures the nonconservation of the energy-momentum tensor, hereafter referred to as the diffusion function, and Eq.~\eqref{originalUGEFE} can be written as
\begin{equation}\label{UGEFE}
R_{\mu\nu}-\frac{1}{2}R g_{\mu\nu}+\Lambda g_{\mu\nu}=\alpha\left(T_{\mu\nu}-Q(x)g_{\mu\nu}\right),
\end{equation}
where the new conservation equation for matter fields is
\begin{equation}\label{Noether}
    \nabla^{\mu}\left(T_{\mu\nu}-Q(x)g_{\mu\nu}\right)=0,
\end{equation}
which is obtained from the Noether theorem, considering that the vector fields $\xi^{\mu}$ that generate the invariance under a restricted group of diffeomorphisms must satisfy the condition $\nabla_{\mu}\xi^{\mu}=0$~\cite{Josset:2016vrq}. Another alternative is to consider the consistency equation
\begin{equation}\label{consistency}
    \nabla^{\mu}\left(R_{\mu\nu}-\frac{1}{2}R g_{\mu\nu}\right)=0,
\end{equation}
which leads to the new conservation equation~\eqref{Noether}~\cite{Ellis:2010uc}.

The set of field equations \eqref{UGEFE} is known as UG, and is an alternative to the GR version \eqref{EFE}, in which $\lambda=\Lambda+Q(x)$ is a variable function, with the cosmological constant $\Lambda$ emerging as an integration constant. The interesting thing about this approach is that the vacuum energy density of quantum fields does not gravitate, since the right-hand side of Eq.~\eqref{UGEFE} is invariant under the simultaneous shift symmetry $T_{\mu\nu}\rightarrow T_{\mu\nu}+\langle\rho\rangle g_{\mu\nu}$ and $Q\rightarrow Q+\langle\rho\rangle$, preventing UG from suffering from the CC problem~\cite{Smolin:2009ti}.

From now on, we consider units in which $\alpha=8\pi G/c^{4}=M_{p}^{2}$, with $M_{p}=2.48\times10^{18}$ GeV.

\section{\label{sec:UnimodularNSC}Non-standard cosmologies in Unimodular Gravity with energy diffusion}
In UG, for a spatially flat FLRW metric with line element
\begin{equation}\label{FLRW}
    ds^2=-dt^{2}+a^{2}(t)\left(dr^{2}+r^{2}d\vartheta^{2}+r^{2}\sin^{2}\left(\vartheta\right) d\varphi^{2}\right),
\end{equation}
and for a matter sector described by the energy-momentum tensor of a perfect fluid,
\begin{equation}\label{energy-momentum}
    T_{\mu\nu}=p\,g_{\mu\nu}+\left(\rho+p\right)u_{\mu}{u_\nu},
\end{equation} 
where $p$, $\rho$, and $u_{\mu}$ are the pressure, energy density, and fluid element four-velocity, respectively, the equations that govern the evolution of the universe take the form
\begin{equation}\label{UGfluids}
    3H^{2}=\frac{\rho+Q}{M_{p}^{2}}+\Lambda,
\end{equation}
\begin{equation}\label{UGpressures}
2\dot{H}+3H^{2}=-\frac{p-Q}{M_{p}^{2}}+\Lambda,
\end{equation}
\begin{equation}\label{UGconservation}
    \dot{\rho}+3H\left(\rho+p\right)=-\dot{Q}.
\end{equation}
As mentioned previously, $Q$ is the diffusion function, which is an arbitrary function that measures the non-conservation of the energy-momentum tensor in UG. It is important to highlight again that these equations reduce to the standard Friedmann equations for GR when $Q=0$.

In general, the energy-momentum tensor does not distinguish between the different fluids that compose the total energy budget of the universe. Thus, we can write the total energy density as the sum of the different matter components, $\rho=\sum_{i}{\rho_{i}}$. Consequently, we can assume that there are different diffusion functions associated with these components, i.e., $Q=\sum_{i}Q_{i}$ \cite{Corral:2020lxt}. In this paper, we are interested in a model in which the early universe is dominated by two fluids: radiation ($\rho_{\gamma}$) and a new field $\phi$ ($\rho_{\phi}$), while the CC ($\Lambda$) and the DM ($\rho_{\chi}$) are negligible in comparison. Although the diffusion function is arbitrary and not necessarily unique, we assume that only the new field $\phi$ experiences energy diffusion. In this sense, we assume that its respective diffusion function is proportional to the energy density of the field, according to the expression $Q=x\rho_{\phi}$. Therefore, Eqs. \eqref{UGfluids} and \eqref{UGconservation} can be written as:
\begin{equation}\label{UGFluid2}
    3H^{2}=\frac{\rho_{\gamma}+\rho_{\phi}+x\rho_{\phi}}{M_{p}^{2}},
\end{equation}
\begin{equation}\label{UGrad}
    \dot{\rho}_{\gamma}+3H\left(\rho_{\gamma}+p_{\gamma}\right)=0,
\end{equation}
\begin{equation}\label{UGfield}
    \dot{\rho}_{\phi}+3H\left(\rho_{\phi}+p_{\phi}\right)=-x\dot{\rho}_{\phi}.
\end{equation}
Note that the sum of Eqs. \eqref{UGrad} and \eqref{UGfield} leads to the conservation equation \eqref{UGconservation} since $\rho=\rho_{\gamma}+\rho_{\phi}$ and $p=p_{\gamma}+p_{\phi}$. Finally, by assuming that the radiation fluid interacts with the field $\phi$, we obtain the governing equations for Unimodular Gravity Non-Standard Cosmologies (UG+NSC), given by
\begin{equation}\label{UGNSCFluid}
    3H^{2}=\frac{\rho_{\gamma}+(x+1)\rho_{\phi}}{M_{p}^{2}},
\end{equation}
\begin{equation}\label{UGNSCfield}
    \dot{\rho}_{\phi}+3H\rho_{\phi}\left(\frac{\omega+1}{1+x}\right)=-\left(\frac{\Gamma_{\phi}}{1+x}\right)\rho_{\phi},
\end{equation}
where the new field $\phi$ is described as a barotropic fluid with an EoS $p_{\phi}=\omega\rho_{\phi}$. Since the conservation equation for the radiation component remains unchanged in this model in comparison to the classical NSC scenario, the radiation energy density can be derived directly from the evolution of the temperature of the universe \eqref{eqtemp}.

Notably, the choice $Q=x\rho_{\phi}$ ensures that when the field $\phi$ fully decays into the SM plasma, the diffusion function becomes negligible, thereby recovering GR and the standard $\Lambda$CDM scenario. Furthermore, this parametrization has been successfully tested with late-time cosmological data~\cite{Corral:2020lxt}.

\begin{figure}
\centering
\includegraphics[width=0.48\textwidth]{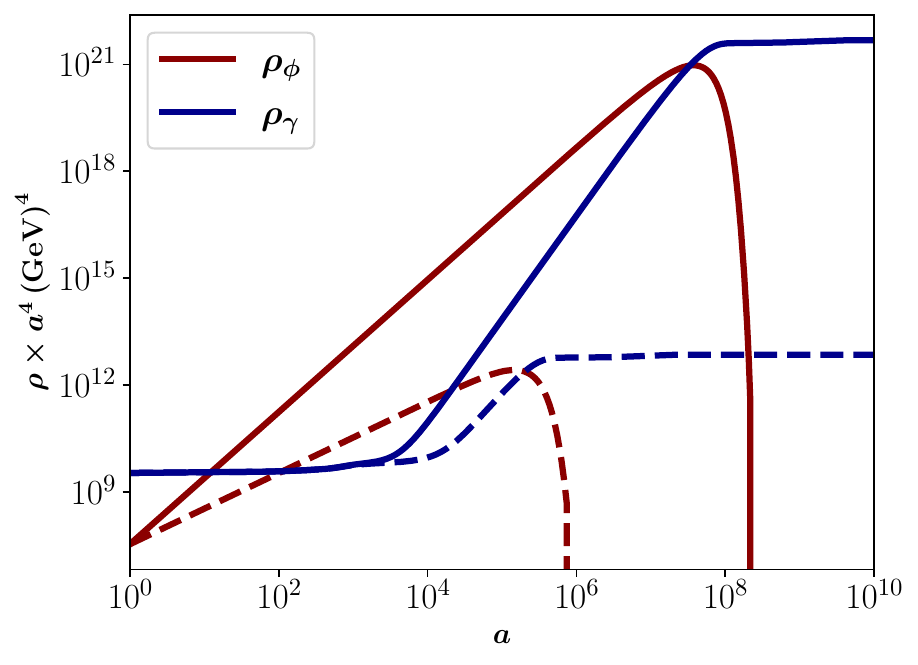}
\caption{Comparison between UG+NSC and NSC scenarios for $\omega=0$, $T_\text{end}=7\times10^{-3}$ GeV, $\kappa=10^{-2}$, $x=0.4$ and $T_i=100$ GeV. The blue and red solid lines correspond to the energy density of radiation and $\phi$ in the UG+NSC scenario, respectively. The blue and red dashed lines correspond to the energy density of radiation and $\phi$ in NSC, respectively.}
\label{Comparacionuninsc}
\end{figure}

Figure \ref{Comparacionuninsc} shows the comparison between the UG+NSC scenario and the NSC case for $x=0.4$, $\omega=0$, $T_\text{end}=7\times10^{-3}$ GeV, $\kappa=10^{-2}$, and $T_i=100$ GeV. The solid and dashed lines represent the evolution of the energy densities of radiation (blue) and $\phi$ (red) in the UG+NSC and NSC scenarios, respectively. This shows that the energy densities scale differently with the scale factor $a$ in UG+NSC, enhancing the entropy injection by $10$ orders of magnitude.

The most important feature of the UG+NSC model is that it slows down both the cosmological dilution of $\phi$ and its effective decay when $x>0$, prolonging its energy presence and modifying the entropy injection into the plasma. The former can be seen in Eq. \eqref{UGNSCfield}, where the dilution term is reduced by a factor $1/(1+x)$ and can be rewritten in terms of an effective barotropic index $\omega_\text{eff}=(\omega-x)/(1+x)$. This means that, for instance, in the case $\omega=0$, the effective equation of state becomes negative, $\omega_\text{eff}<0$, mimicking a quintessence-like behavior. However, this does not arise from a fundamental scalar field, but from the modified dynamics induced by diffusion in UG. For a given value of $\omega$, different values of $\omega_\text{eff}$ can be obtained depending on the strength of the diffusion parameter. Also, from the decay term, an effective decay rate for $\phi$ can be identified as $\Gamma_{\phi_\text{eff}}=\Gamma_\phi/(1+x)$, which is lower than the decay rate in the NSC scenario, i.e., the $\phi$ particle decays more slowly in this scenario. On the other hand, for $x<0$, both effects are enhanced, leading to a faster depletion of $\rho_\phi$ and reducing its cosmological impact. For these reasons, we restrict our analysis to $x>0$, where the model consistently enhances the lifetime and cosmological relevance of $\phi$.

\subsection{\label{subsec:Comparison} Comparison between scenarios}
\begin{figure}
\centering
\includegraphics[width=0.48\textwidth]{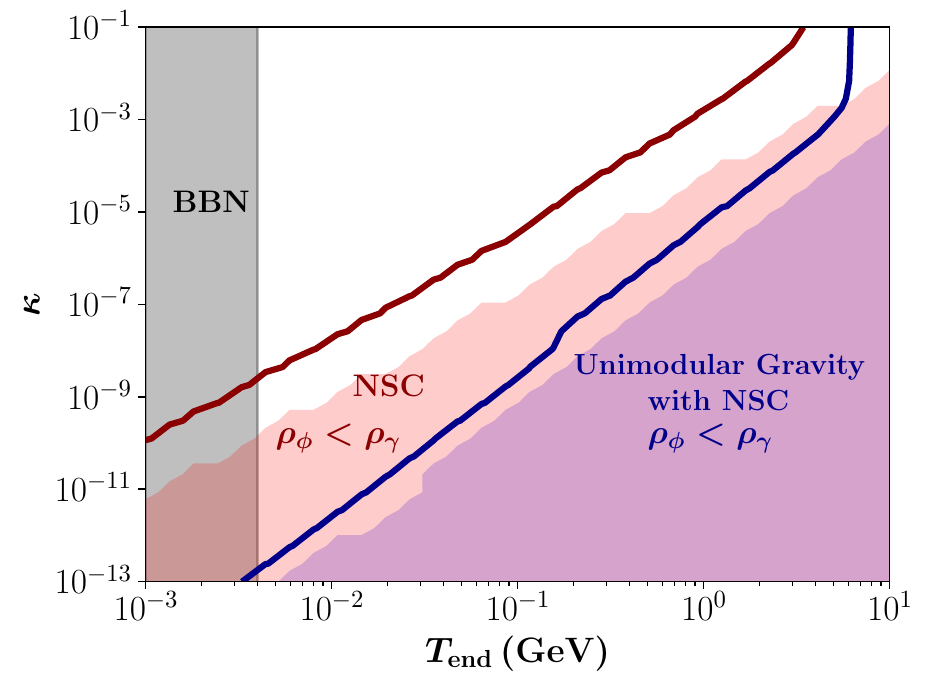}
\caption{$T_\text{end}$-$\kappa$ parameter space for WIMP DM with $m_\chi=100$ GeV, $\langle\sigma v\rangle=10^{-11}$ GeV$^{-2}$, $\omega=-2/5$, and $x=0.6$. The blue and red lines correspond to the parameters that reproduce the current DM relic density in UG+NSC and standard NSC scenario, respectively. The grey region corresponds to the BBN epoch, which starts at $T_\text{BBN}\sim4\times10^{-3}$ GeV. The blue and red regions (UG+NSC and NSC scenario, respectively) correspond to the parameters in which the energy of $\phi$ is always lower than that of radiation.}
\label{ComparacionUGNSCw-25}
\end{figure}

The new cosmological scenario introduced by UG, incorporating a field $\phi$ in the early Universe, opens up new regions of parameter space in the search for DM candidates. Figure \ref{ComparacionUGNSCw-25} shows a comparison between the NSC (red line) and UG+NSC (blue line) scenarios for a WIMP DM candidate with $m_\chi = 100$ GeV, $\langle \sigma v \rangle = 10^{-11}$ GeV$^{-2}$, $\omega = -2/5$, and $x = 0.6$. The red and blue lines correspond to the parameter values that reproduce the observed DM relic density, while the shaded regions indicate where $\rho_\phi < \rho_\gamma$ at all times. The grey region corresponds to the BBN epoch, starting at a temperature of approximately $4 \times 10^{-3}$ GeV. In this case, the fluid behaves as one with an effective barotropic index $\omega=-5/8$ in the UG+NSC scenario, even though its intrinsic equation of state is $\omega=-2/5$. From the figure, we observe that this implies that the UG+NSC scenario allows smaller values of $\kappa$ within a similar range of $T_\text{end}$ (spanning approximately four orders of magnitude). This is due to the slower dilution and decay of $\phi$, so its energy density remains relevant for a longer period, enhancing its cosmological impact. As a result, smaller values of $\kappa$ are sufficient to produce similar non-standard cosmological effects. However, both scenarios exhibit a similar dependence in the $\kappa$–$T_\text{end}$ parameter space, with comparable shapes of the curves but different parameter values.

From now on, in all figures of this section, the blue and red shaded regions indicate the parameter space where $\rho_\phi < \rho_\gamma$, corresponding to times when the energy density of the field $\phi$ never dominates over radiation. The blue color corresponds to the UG+NSC scenario, while red represents the standard NSC scenario.

\begin{figure}
\centering
\includegraphics[width=0.48\textwidth]{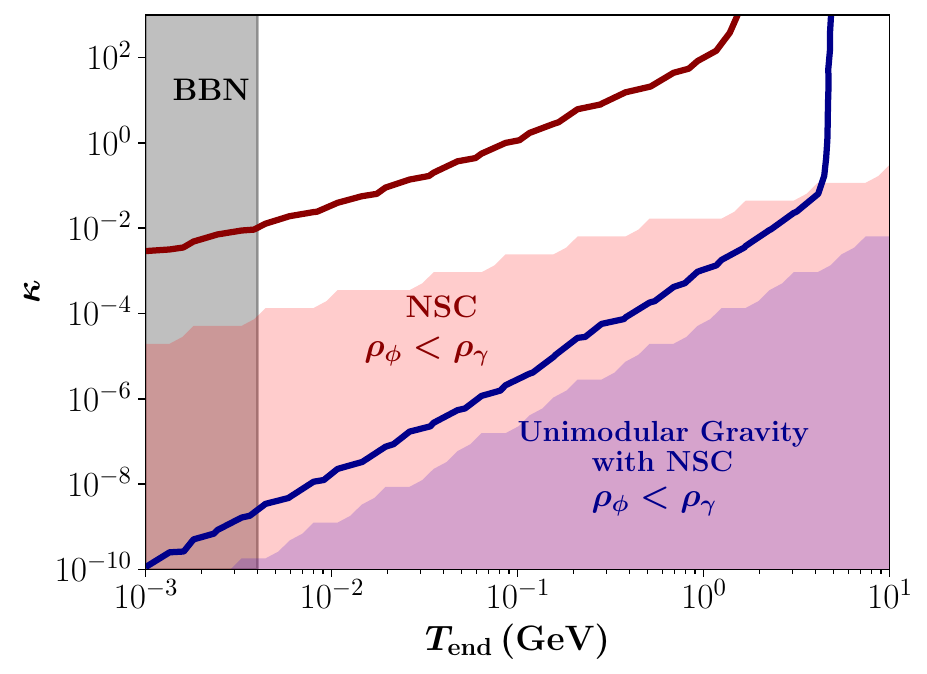}
\caption{$T_\text{end}$–$\kappa$ parameter space for WIMP DM with $m_\chi = 100$ GeV, $\langle\sigma v\rangle = 10^{-11}$ GeV$^{-2}$, $\omega = 0$, and $x = 0.6$. The blue and red lines correspond to the parameters that reproduce the current DM relic density in the UG+NSC and standard NSC scenarios, respectively. The grey region corresponds to the BBN epoch, which starts at $T_\text{BBN} \sim 4 \times 10^{-3}$ GeV. The blue and red shaded regions (for UG+NSC and NSC scenarios, respectively) indicate the parameter space in which the energy density of $\phi$ is always lower than that of radiation.}
\label{ComparacionUGNSCw0}
\end{figure}

The case $\omega = 0$ is shown in Figure \ref{ComparacionUGNSCw0}, which compares the NSC and UG+NSC scenarios for a WIMP DM candidate with $m_\chi = 100\,\text{GeV}$, $\langle \sigma v \rangle = 10^{-11}\,\text{GeV}^{-2}$, and $x = 0.6$. The red and blue lines correspond to the parameter values that reproduce the observed DM relic density in NSC and UG+NSC, respectively. The grey region corresponds to the BBN epoch, starting at $T_\text{BBN} \sim 4 \times 10^{-3}\,\text{GeV}$. From this figure, we observe that in the UG+NSC scenario, the allowed parameter space is shifted toward lower values of $\kappa$ for the same range of $T_\text{end}$ compared to the standard NSC case. This shift amounts to approximately seven orders of magnitude. In addition, the slope in this scenario is steeper than in the standard NSC case, although the qualitative behavior remains similar at large $T_\text{end}$, where a $\kappa$-independent region appears.

The change in slope between the two scenarios can be understood in terms of the effective equation of state. For $x = 0.6$, even a pressureless fluid ($\omega = 0$) develops a negative effective barotropic index, $\omega_{\text{eff}}<0$, resulting in a slower dilution of $\rho_\phi$. This enhances the persistence of the field and explains the shift in the $\kappa$-$T_{\text{end}}$ parameter space, where smaller values of $\kappa$ yield comparable to NSC effects.

\begin{figure}
\centering
\includegraphics[width=0.48\textwidth]{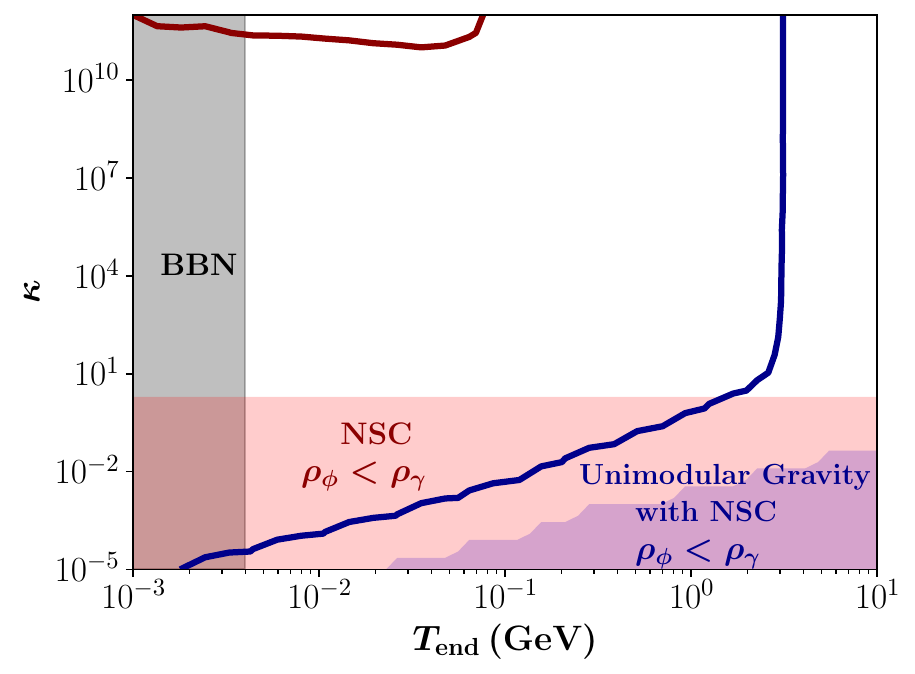}
\caption{$T_\text{end}$-$\kappa$ parameter space for WIMP DM with $m_\chi=100$ GeV, $\langle\sigma v\rangle=10^{-11}$ GeV$^{-2}$, $\omega=2/5$, and $x=0.6$. The blue and red lines correspond to the parameters that reproduce the current DM relic density in UG+NSC and standard NSC scenario, respectively. The grey region corresponds to the BBN epoch, which starts at $T_\text{BBN}\sim4\times10^{-3}$ GeV. The blue and red regions (UG+NSC and NSC scenario, respectively) indicate the parameter space in which the energy of $\phi$ is always lower than that of radiation.}
\label{ComparacionUGNSCw25}
\end{figure}

The comparison between NSC (red line) and UG+NSC (blue line) for a WIMP DM candidate with $m_\chi = 100\,\text{GeV}$, $\langle \sigma v \rangle = 10^{-11}\,\text{GeV}^{-2}$, $\omega = 2/5$, and $x = 0.6$ is shown in Figure \ref{ComparacionUGNSCw25}. The grey region corresponds to the BBN epoch, starting at a temperature of approximately $4 \times 10^{-3}\,\text{GeV}$. In this case, the region of parameter space that reproduces the observed DM relic density differs significantly from the previous scenarios, with the gap between the two models increasing to about 16 orders of magnitude at lower values of $T_\text{end}$. The dependence on the $T_\text{end}$--$\kappa$ parameters in the UG+NSC scenario remains qualitatively similar to the cases with $\omega = -2/5$ and $\omega = 0$. This behavior can be understood in terms of the effective barotropic index: for this case, $\omega_\text{eff} = -1/8$, in contrast to the intrinsic value $\omega = 2/5$, which in the standard NSC scenario would correspond to a fluid that dilutes faster than radiation. 
However, increasing $\omega$ amplifies the difference between the two scenarios, leading to larger separations in parameter space and allowing smaller values of $\kappa$ to reproduce the observed DM relic density, due to the faster dilution of the energy density in the standard NSC case compared to its effective behavior in UG+NSC.

\begin{figure}
\centering
\includegraphics[width=0.48\textwidth]{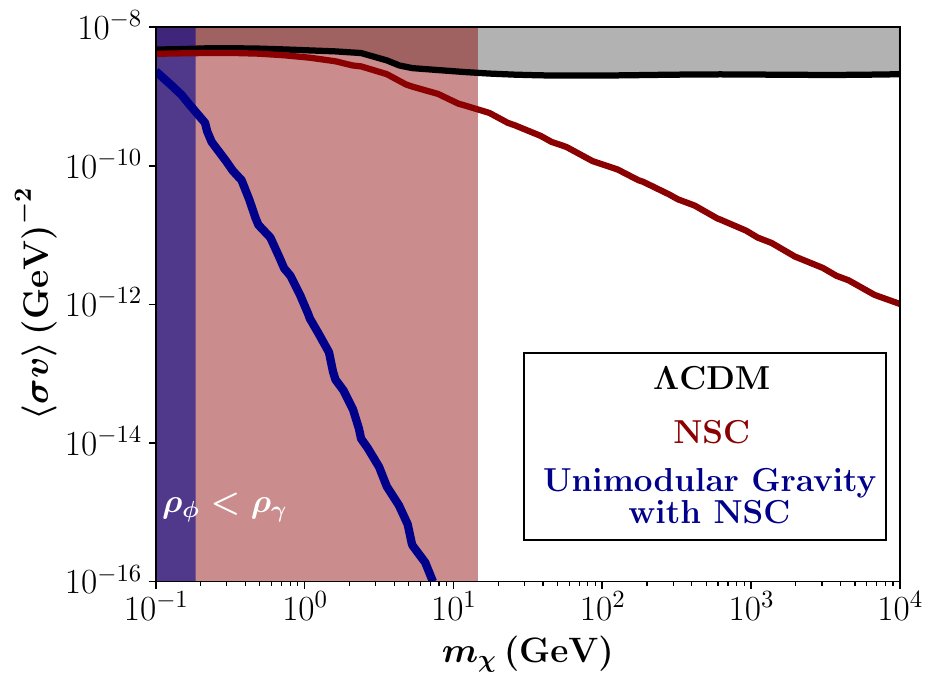}
\caption{$m_\chi$–$\langle\sigma v\rangle$ parameter space for WIMP DM with $\kappa = 10^{-3}$, $T_\text{end} = 7 \times 10^{-3}$ GeV, $\omega = 0$, and $x = 0.6$. The blue, red, and black lines correspond to the parameters that reproduce the current DM relic density in the UG+NSC, NSC, and $\Lambda$CDM scenarios, respectively. The grey region corresponds to the excluded parameters, which are also restricted in the UG scenario. The blue and red shaded regions correspond to the parameters for which the energy density of $\phi$ is always lower than that of radiation, for the UG+NSC and NSC scenarios, respectively.}
\label{ComparacionNSCUG}
\end{figure}

We now turn to the comparison in the DM parameter space, $(m_\chi, \langle \sigma v \rangle)$, between the UG+NSC scenario and the standard NSC cosmology, as shown in Figure \ref{ComparacionNSCUG}. For a WIMP DM candidate with $\kappa = 10^{-3}$, $T_\text{end} = 7 \times 10^{-3}\, \text{GeV}$, $\omega = 0$, and $x = 0.6$, the blue, red, and black lines indicate the parameter values that reproduce the observed DM relic density in UG+NSC, NSC, and $\Lambda$CDM, respectively. The grey region corresponds to excluded parameters for WIMP production in $\Lambda$CDM, which are also constrained in this scenario. Notably, in the UG+NSC scenario, the required values of $\langle \sigma v \rangle$ differ significantly from those in the standard NSC case. In particular, the steeper slope of the curve arises from the modified effective dynamics: the slower dilution associated with $\omega_\text{eff}$, together with the delayed decay governed by $\Gamma_{\phi_\text{eff}}$. As a result, the energy density of $\phi$ remains relevant for a longer period, allowing smaller values of $\langle \sigma v \rangle$ for lighter DM candidates and opening new regions of parameter space compatible with the observed relic density.

In the following subsections, we focus on the impact of varying the model and DM parameters in the UG+NSC scenario to reproduce the observed DM relic density for WIMP DM candidates, with the Real Singlet Scalar DM candidate as a case of study.

\subsection{\label{subsec:WIMPParameters} Parameter spaces for the WIMP DM}
We first analyze WIMP DM by fixing the particle properties to study how the model variables $(T_\text{end}-\kappa)$ and the diffusion factor $(x)$ change for specific values of $m_\chi$ and $\langle\sigma v\rangle$.

\begin{figure}
\centering
\includegraphics[width=0.48\textwidth]{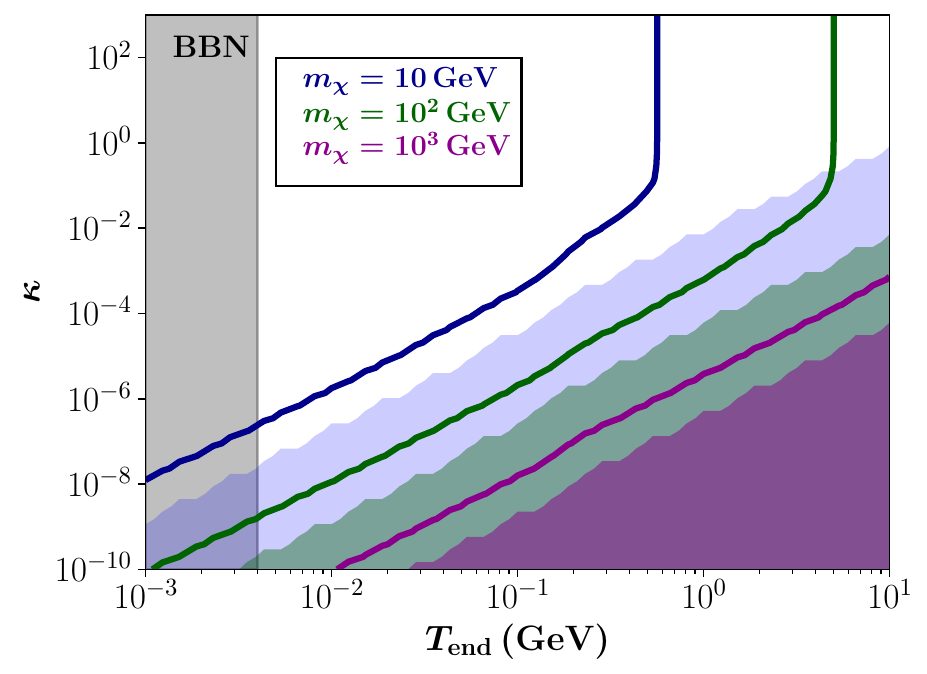}
\caption{$T_\text{end}$-$\kappa$ parameter space for WIMP DM with $\langle\sigma v\rangle=10^{-11}$ GeV$^{-2}$, $x=0.6$ and $\omega=0$. The blue, green, and magenta lines correspond to the parameters that reproduce the current DM relic density in  UG+NSC for DM mass $m_\chi=10$, $m_\chi=10^2$, and $m_\chi=10^3$ GeV, respectively. The grey region corresponds to the BBN epoch, which starts at $T_\text{BBN}\sim4\times10^{-3}$ GeV. The blue, green, and magenta regions correspond to the parameters in which the energy of $\phi$ is always lower than that of radiation when $m_\chi=10$, $m_\chi=10^2$, and $m_\chi=10^3$ GeV, respectively.}
\label{varmsvcte}
\end{figure}

Figure \ref{varmsvcte} shows the $T_\text{end}$–$\kappa$ parameter space for different WIMP DM masses, with $\langle\sigma v\rangle = 10^{-11}$ GeV$^{-2}$, $x = 0.6$, and $\omega = 0$. The blue, green, and magenta lines indicate the parameters that reproduce the observed DM relic density in the UG+NSC scenario for $m_\chi = 10$, $10^2$, and $10^3$ GeV, respectively. The grey region corresponds to the BBN epoch, which starts at $T_\text{BBN} \sim 4\times10^{-3}\;\text{GeV}$. Shaded regions correspond to the energy of $\phi$ being always lower than that of radiation, with the color of the shaded areas reflecting the different DM masses analyzed.
As the DM mass increases, the allowed parameter space shifts downward and to the right, extending to lower values of $\kappa$ and higher values of $T_\text{end}$. This behavior can be understood from the freeze-out timing. Lighter DM particles decouple later, typically approaching the regime where $\phi$ decays become significant (RIII). In this case, the DM abundance is less affected by subsequent entropy injection, requiring larger values of $\kappa$ to reproduce the observed relic density.
On the other hand, heavier DM particles freeze out earlier, generally during the radiation-dominated phase (RI), and thus experience a stronger dilution due to the later decay of $\phi$. As a consequence, smaller values of $\kappa$ are sufficient to match the current DM relic density.

\begin{figure}
\centering
\includegraphics[width=0.48\textwidth]{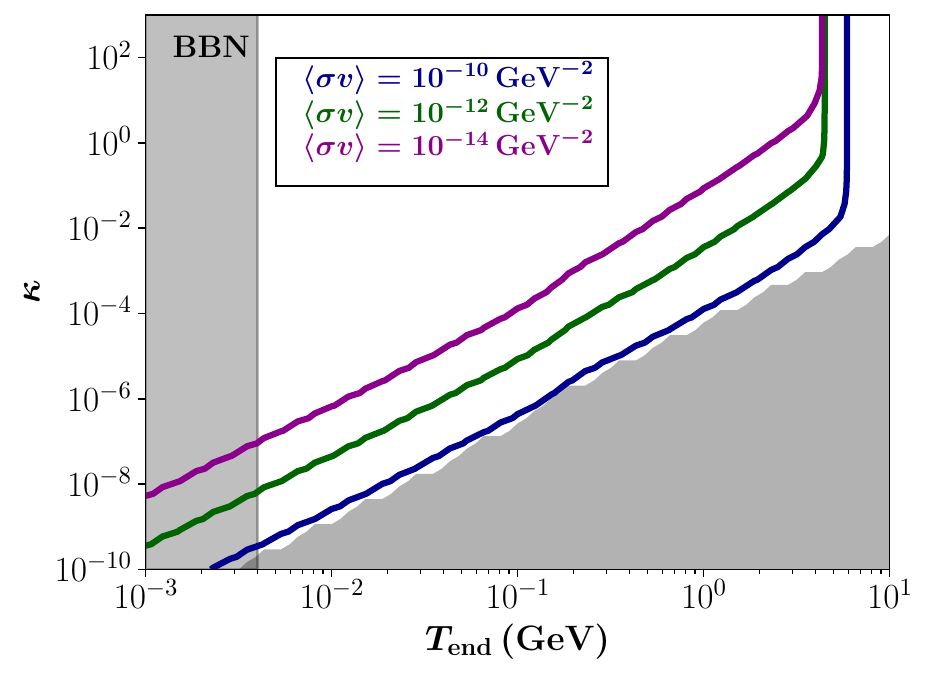}
\caption{$T_\text{end}$-$\kappa$ parameter space for $\langle\sigma v\rangle$, with WIMP DM $m_\chi=100$ GeV, $x=0.6$ and $\omega=0$. The blue, green, and magenta lines correspond to the parameters that reproduce the current DM relic density in UG+NSC for $\langle\sigma v\rangle=10^{-10}$, $\langle\sigma v\rangle=10^{-12}$, and $\langle\sigma v\rangle=10^{-14}$ GeV$^{-2}$, respectively. The grey region corresponds to the BBN epoch, which starts at $T_\text{BBN}\sim4\times10^{-3}$ GeV. The lower grey region corresponds to the parameters in which the energy of $\phi$ is always lower than that of radiation for the three values of $\langle\sigma v\rangle$.}
\label{varsvmcte}
\end{figure}

The $T_\text{end}$–$\kappa$ space that reproduces the DM relic density for a fixed WIMP mass is presented in Figure \ref{varsvmcte}. In this figure, the total thermally averaged annihilation cross-section is varied for $m_\chi = 100$ GeV, $x = 0.6$, and $\omega = 0$. The blue, green, and magenta lines correspond to the values of $\langle\sigma v\rangle = 10^{-10}$, $10^{-12}$, and $10^{-14}$ GeV$^{-2}$, respectively, that reproduce the observed DM relic density in the UG+NSC scenario. The grey region indicates the BBN epoch, which begins at $T_\text{BBN} \sim 4\times10^{-3}$ GeV, while the lower shaded region represents the range where the energy of $\phi$ is always below that of radiation; this region is the same for all three cross-section values. It can be seen that smaller values of the thermally averaged cross-section correspond to higher values of $\kappa$ within the same range of $T_\text{end}$, with no significant modification of the allowed $T_\text{end}$ interval. Consequently, increasing $\langle\sigma v\rangle$ shifts the allowed region downward in the $T_\text{end}$–$\kappa$ plane. This behavior is expected, as larger values of $\langle\sigma v\rangle$ enhance DM annihilation, reducing the final relic abundance, so smaller values of $\kappa$ are required to reproduce the observed DM relic density. 

\begin{figure}
\centering
\includegraphics[width=0.48\textwidth]{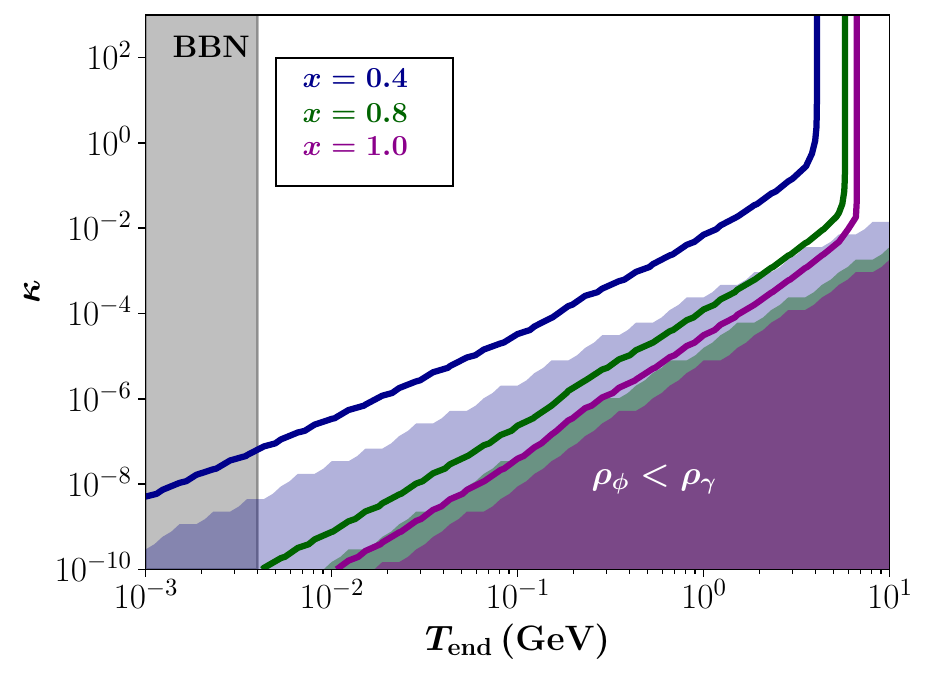}
\caption{$T_\text{end}$-$\kappa$ parameter space for WIMP DM with $m_\chi=100$ GeV, $\langle\sigma v\rangle=10^{-11}$ GeV$^{-2}$, and $\omega=0$. The blue, green, and magenta lines correspond to the parameters that reproduce the current DM relic density in  UG+NSC for values $x=0.4$, $x=0.8$, and $x=1.0$, respectively. The grey region corresponds to the BBN epoch, which starts at $T_\text{BBN}\sim4\times10^{-3}$ GeV. The blue, green, and magenta regions correspond to the parameters in which the energy of $\phi$ is always lower than that of radiation when $x=0.4$, $x=0.8$, and $x=1.0$, respectively.}
\label{ComparacionUGx}
\end{figure}

Figure \ref{ComparacionUGx} shows the $T_\text{end}$–$\kappa$ space when the diffusion parameter $x$ is varied for WIMP DM with $m_\chi = 100$ GeV, $\langle\sigma v\rangle = 10^{-11}$ GeV$^{-2}$, and $\omega = 0$. The blue, green, and magenta lines correspond to the values of $x = 0.4$, $0.8$, and $1.0$, respectively, that reproduce the observed DM relic density in the UG+NSC scenario. The grey region represents the BBN epoch, which begins at $T_\text{BBN} \sim 4\times10^{-3}$ GeV. Shaded regions indicate where the energy of $\phi$ is always lower than that of radiation, with the color of each region corresponding to the associated $x$ value. Increasing $x$ affects the slope of the curves that reproduce the observed DM relic density: larger values of $x$ lead to steeper slopes, reflecting the decrease in the effective equation of state $\omega_\text{eff}$. This behavior is also observed in Figure \ref{ComparacionUGNSCw0}, where the case $x=0.6$ exhibits a more pronounced slope compared to $x=0$, corresponding to the standard NSC scenario. Additionally, increasing $x$ produces a slight rightward shift of the curves in the $T_\text{end}$–$\kappa$ plane. In the limit $x \rightarrow 0$, the standard NSC behavior is recovered, characterized by a shallower slope and a leftward shift of the allowed region. Conversely, $x \rightarrow 1$ approaches the limiting value $\omega_\text{eff} \rightarrow -0.5$ for an intrinsic equation of state $\omega = 0$.

We now take the opposite approach, fixing the model parameters to constrain the WIMP DM candidate parameter space $\left(m_\chi-\langle\sigma v\rangle\right)$ in order to reproduce the observed DM relic density.

\begin{figure}
\centering
\includegraphics[width=0.48\textwidth]{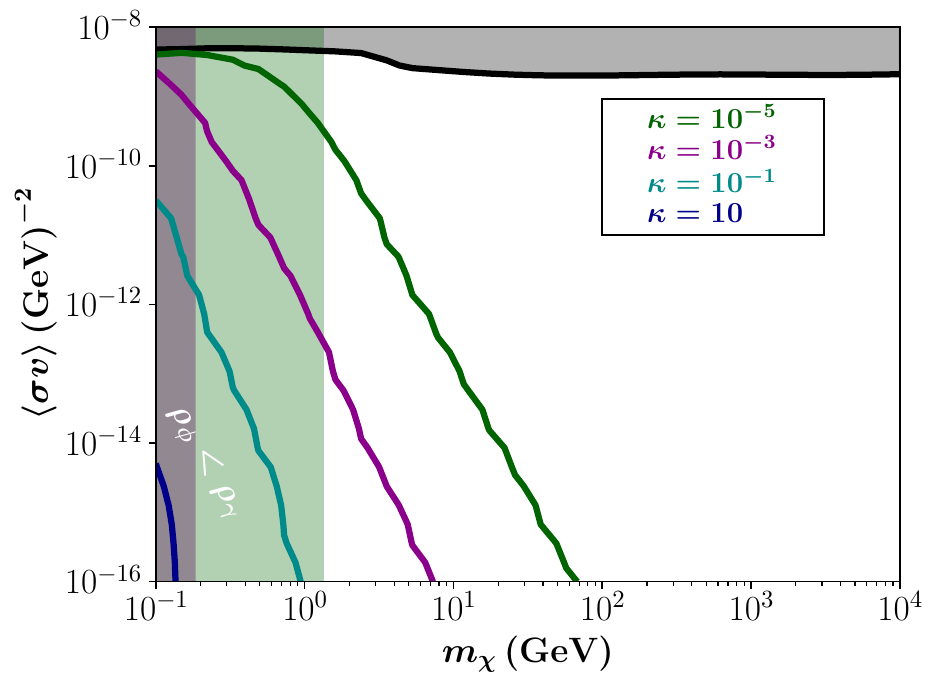}
\caption{$m_\chi$-$\langle\sigma v\rangle$ parameter space for WIMP DM with $T_\text{end}=7\times10^{-3}$ GeV, $x=0.6$, and $\omega=0$. The green, magenta, cyan, and blue lines correspond to the parameters that reproduce the current DM relic density in UG+NSC for $\kappa=10^{-5}$, $10^{-3}$, $10^{-1}$, and $10$, respectively. The black line corresponds to the parameters that reproduce the DM relic density in $\Lambda$CDM $(\langle\sigma v\rangle_0=\text{few}\times10^{-9}$ GeV$^{-2}$). The grey region corresponds to the excluded parameters in $\Lambda$CDM, which in this case is also restricted for the UG+NSC scenario. The green, cyan, magenta, and blue regions correspond to the parameters in which the energy of $\phi$ is always lower than that of radiation, for $\kappa=10^{-5}$, $10^{-3}$, $10^{-1}$, and $10$, respectively.}
\label{variacionkappadm}
\end{figure}

Figure \ref{variacionkappadm} shows the $m_\chi$–$\langle\sigma v\rangle$ plane for WIMP DM in the UG+NSC scenario, with $T_\text{end} = 7\times10^{-3}$ GeV, $x = 0.6$, and $\omega = 0$. The green, magenta, cyan, and blue lines correspond to $\kappa = 10^{-5}$, $10^{-3}$, $10^{-1}$, and $10$, respectively, reproducing the observed DM relic density. The black and grey regions indicate the parameters that reproduce and overproduce the DM relic density in the standard $\Lambda$CDM scenario. Magenta and green shaded areas mark where $\rho_\phi$ is always below $\rho_\gamma$ for $\kappa = 10^{-5}$ and $10^{-3}$, respectively. Increasing $\kappa$ shifts the curves to the left in the $\langle \sigma v \rangle$--$m_\chi$ plane, allowing smaller DM masses to reproduce the observed relic density for the same range of interaction strength. This behavior reflects the modified cosmological evolution, which changes the relation between the DM mass and the thermally averaged cross-section required to match the observed relic density. Heavier DM particles, which freeze out earlier, are generally more sensitive to the subsequent dilution effects, while lighter particles decouple later and are less affected by them.

\begin{figure}
\centering
\includegraphics[width=0.48\textwidth]{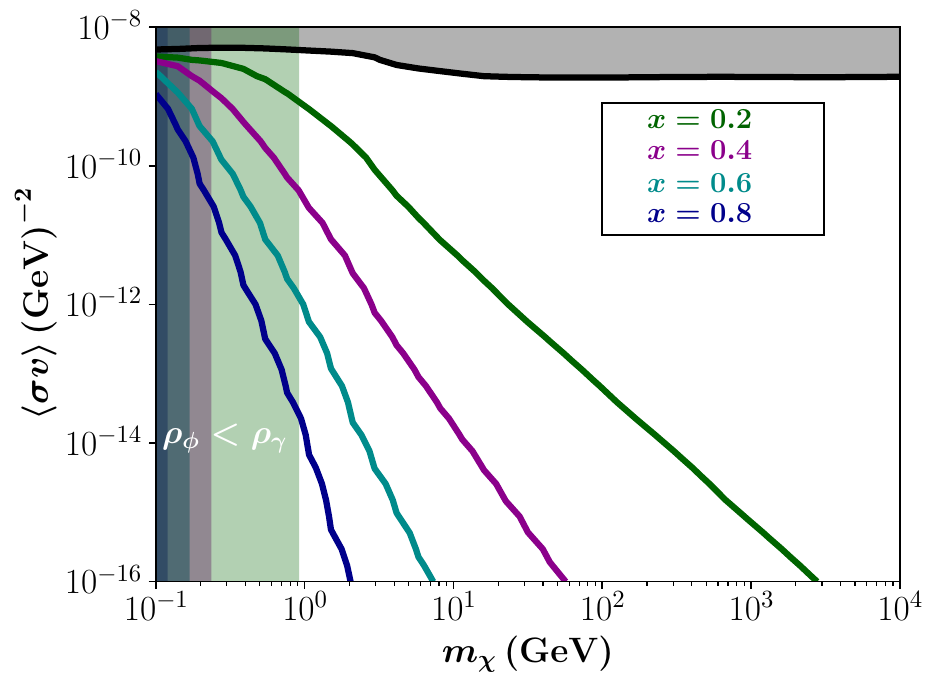}
\caption{$m_\chi$-$\langle\sigma v\rangle$ parameter space for WIMP DM with $\kappa=10^{-3}$, $T_\text{end}=7\times10^{-3}$ GeV, and $\omega=0$. The green, magenta, cyan, and blue lines correspond to the parameters that reproduce the current DM relic density in UG+NSC for $x=0.2$, $0.4$, $0.6$, and $0.8$, respectively. The black line corresponds to the parameters that reproduce the DM relic density in $\Lambda$CDM $(\langle\sigma v\rangle_0=\text{few}\times10^{-9}$ GeV$^{-2}$). The grey region corresponds to the excluded parameters in $\Lambda$CDM, which in this case is also restricted for the UG+NSC scenario. The green, cyan, magenta, and blue regions correspond to the parameters in which the energy of $\phi$ is always lower than that of radiation, for $x=0.2$, $0.4$, $0.6$, and $0.8$, respectively.}
\label{Variacionxdm}
\end{figure}

The impact of varying the diffusion parameter $x$ on the WIMP DM parameter space is shown in Figure \ref{Variacionxdm}, for $\kappa = 10^{-3}$, $T_\text{end} = 7\times10^{-3}$ GeV, and $\omega = 0$. The green, magenta, cyan, and blue lines correspond to $x = 0.2$, $0.4$, $0.6$, and $0.8$, respectively, reproducing the observed DM relic density in the UG+NSC scenario. The black line and grey region indicate the allowed and excluded regions in the standard $\Lambda$CDM scenario; note that the $\Lambda$CDM excluded region is still viable in the UG scenario. Shaded areas of matching colors mark where the energy of $\phi$ is always below that of radiation for each $x$ value. As $x$ increases, the curves shift downward and become steeper, reducing the maximum DM mass compatible with the observed relic density. Equivalently, larger values of $x$ allow smaller $\langle\sigma v\rangle$ at lower DM masses. This behavior originates from $\omega_\text{eff}<\omega$, which alters the cosmological evolution of the $\phi$ field, leading to a slower dilution compared to the standard pressureless case.

\begin{figure}
\centering
\includegraphics[width=0.48\textwidth]{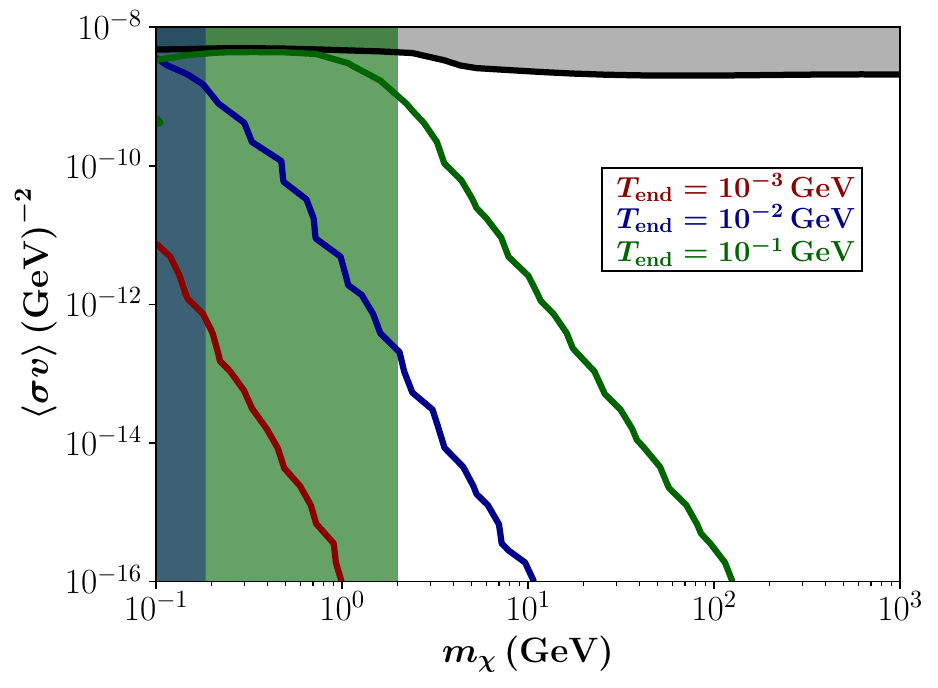}
\caption{$m_\chi$-$\langle\sigma v\rangle$ parameter space for WIMP DM with $\kappa=10^{-3}$, $\omega=0$, and $x=0.6$. The red, blue, and green lines correspond to the parameters that reproduce the current DM relic density in UG+NSC for $T_\text{end}=10^{-3}$, $10^{-2}$, and $10^{-1}$ GeV, respectively. The black line corresponds to the parameters that reproduce the DM relic density in $\Lambda$CDM $(\langle\sigma v\rangle_0=\text{few}\times10^{-9}$ GeV$^{-2}$). The grey region corresponds to the excluded parameters in $\Lambda$CDM, which in this case is also restricted for the UG+NSC scenario. The blue and green regions correspond to the parameters in which the energy of $\phi$ is always lower than that of radiation, for $T_\text{end}=10^{-2}$ and $10^{-1}$ GeV, respectively.}
\label{variaciontendDMk1e-3w0}
\end{figure}

The effect of varying $T_\text{end}$ on the WIMP DM parameter space is shown in Figure \ref{variaciontendDMk1e-3w0}, for $\kappa = 10^{-3}$, $\omega = 0$, and $x = 0.6$. The red, blue, and green lines correspond to $T_\text{end} = 10^{-3}$, $10^{-2}$, and $10^{-1}$ GeV, respectively, reproducing the observed DM relic density in the UG+NSC scenario. Blue and green shaded areas indicate where $\rho_\phi < \rho_\gamma$. As $T_\text{end}$ increases, the curves shift to the right, allowing higher DM masses to reproduce the observed relic density. A larger $T_\text{end}$ corresponds to an earlier decay of the $\phi$ field, reducing the amount of entropy injection after freeze-out. The resulting shift reflects the interplay between the modified expansion history, the freeze-out process, and the reduced dilution, which together alter the relation between $m_\chi$ and $\langle\sigma v\rangle$.

\begin{figure}
\centering
\includegraphics[width=0.48\textwidth]{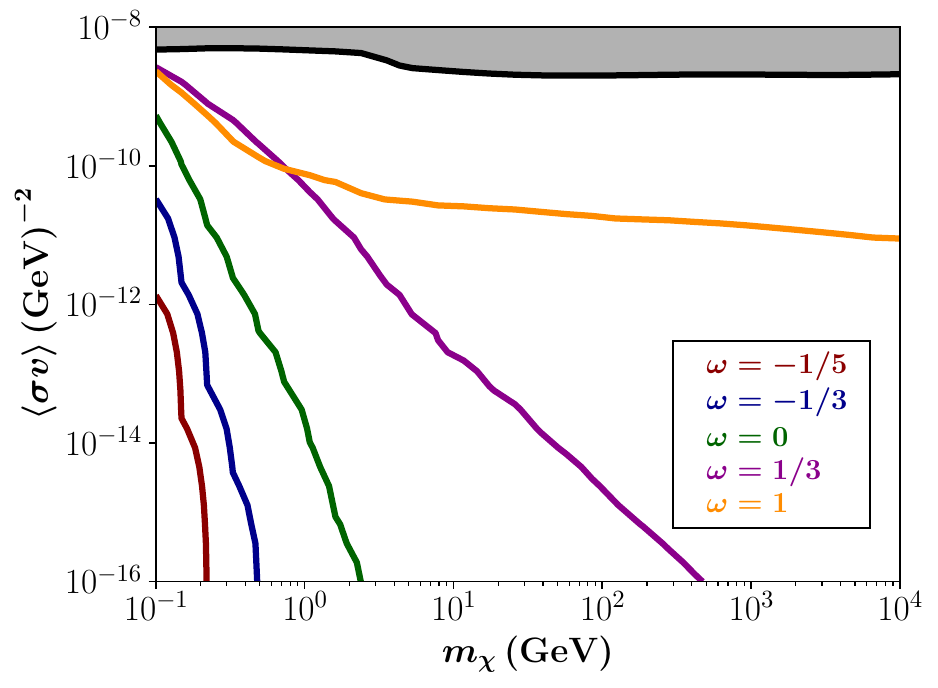}
\caption{$m_\chi$-$\langle\sigma v\rangle$ parameter space for WIMP DM for $T_\text{end}=7\times10^{-3}$ GeV, and $x=0.6$. The red, blue, green, magenta, and orange lines correspond to the parameters that reproduce the current DM relic density in UG+NSC for $\omega=-1/3$, $-1/5$, $0$, and $1/3$ with $\kappa=10^{-2}$, and $\omega=1$ with $\kappa=10^4$, respectively. The black line corresponds to the parameters that reproduce the DM relic density in $\Lambda$CDM $(\langle\sigma v\rangle_0=\text{few}\times10^{-9}$ GeV$^{-2}$). The grey region corresponds to the excluded parameters in $\Lambda$CDM, which in this case is also restricted for the UG+NSC scenario. The regions in which $\rho_\phi$ is always less than $\rho_\gamma$ are still present for $\kappa=10^{-2}$ to lower values of $m_\chi$ than those presented in the plot.}
\label{variacionomegawimp}
\end{figure}

Figure \ref{variacionomegawimp} shows the DM parameter space that reproduces the observed relic density for different values of the barotropic index $\omega$ of the field $\phi$, with $T_\text{end} = 7 \times 10^{-3}\;\text{GeV}$ and $x = 0.6$ in the UG+NSC scenario. The red, blue, green, magenta, and orange lines correspond to $\omega = -1/3$, $-1/5$, $0$, $1/3$ with $\kappa = 10^{-2}$, and $\omega = 1$ with $\kappa = 10^4$, respectively. Increasing $\omega$ with $\kappa = 10^{-2}$ allows the curves to reach higher DM masses for the same $\langle\sigma v\rangle$, not by shifting the curves but through a change in their slope. As $\omega$ increases, the energy density of $\phi$ dilutes faster than radiation. For $\omega = 1$, $\kappa = 10^{-2}$ nearly reproduces the $\Lambda$CDM parameter space, which is why a higher $\kappa$ is required to compensate for the faster dilution of $\phi$, resulting in a larger entropy injection. In this case ($\omega = 1$, $\kappa = 10^4$), the DM parameter space shifts downward and changes its slope compared to the $\Lambda$CDM parameters, as previously observed in Figure \ref{variacionkappadm}.

The results presented above are entirely general, in the sense that they rely on the DM mass $m_\chi$ and the thermally averaged annihilation cross section $\langle \sigma v\rangle$, independently of the underlying particle physics realization of the WIMP. This generality is a strength for the NSC framework in diffusive UG, since the same qualitative conclusions apply to any WIMP candidate, regardless of the specific model that generates it. However, in order to confront our results with concrete experimental constraints and to illustrate how the modified cosmological evolution translates into bounds on physical couplings, it is instructive to show the impact of our results on a concrete model. Among the minimal WIMP realizations available in the literature, we adopt the Real Singlet Scalar extension of the SM as a benchmark model, given its simplicity, its well-known relic density and direct detection phenomenology, and the fact that it depends on a single free coupling besides $m_\chi$, which allows for a direct and unambiguous mapping between our general $(m_\chi,\langle\sigma v\rangle)$ results and a testable particle physics parameter space.

\subsection{\label{subsec:RSSParameters} Real singlet DM parameter space}
We now apply the previous general results to a well-motivated and minimal realization of WIMP DM, in which the SM is extended by a real scalar field $\chi$, singlet under the SM gauge group $SU(3)_c \times SU(2)_L \times U(1)_Y$, and stabilized against decay into SM particles by an exact $\mathbb{Z}_2$ symmetry under which $\chi \to -\chi$ while all SM fields remain even. The most general renormalizable Lagrangian for this extension is given by
\begin{equation}
    \mathscr{L} \supset \frac{1}{2}(\partial_\mu \chi)(\partial^\mu \chi) - \frac{1}{2}\mu_\chi^2 \chi^2 - \frac{1}{4}\lambda_\chi \chi^4 - \frac{1}{2}\lambda_{HS}\,\chi^2 \left(H^\dagger H\right),
    \label{eq:RSSLagrangian}
\end{equation}
where $H$ is the SM Higgs doublet, $\mu_\chi$ and $\lambda_\chi$ are the bare mass and self-coupling of $\chi$, and $\lambda_{HS}$ is the Higgs-portal coupling connecting the dark and visible sectors. After electroweak symmetry breaking, $H^\dagger H \supset \frac{1}{2}(v+h)^2$, so that the physical DM mass reads $m_\chi^2 = \mu_\chi^2 + \tfrac{1}{2}\lambda_{HS}v^2$, with $v\simeq246$ GeV the Higgs vacuum expectation value. The $\mathbb{Z}_2$ symmetry forbids a $\chi$-$H$ mixing term and a linear term in $\chi$, ensuring that $\chi$ is absolutely stable and does not develop a vacuum expectation value, as required for a viable DM candidate.

In this minimal setup, the phenomenology of $\chi$ is entirely controlled by only two parameters, $m_\chi$ and $\lambda_{HS}$, since $\lambda_\chi$ does not enter the DM production or detection observables at tree level. We restrict our analysis to $m_\chi\gtrsim200$~GeV, i.e., above the top-quark threshold and sufficiently far from the Higgs resonance at $m_\chi\simeq m_h/2$ and the $W$, $Z$, $h$, and $t\bar t$ production thresholds. In this regime, all relevant SM annihilation channels ($\chi\chi\to h^*\to b\bar b, WW, ZZ, t\bar t$ through s-channel Higgs exchange, together with the direct $\chi\chi\to hh$ process) are kinematically open, and the canonical thermally averaged cross section \cite{Gondolo:1990dk} reduces, to excellent approximation, to its velocity independent value evaluated at threshold, $\langle\sigma v\rangle \simeq \sigma v_\text{rel}\big|_{s=4m_\chi^2}$ \cite{Cirelli:2024ssz}, which depends only on $(m_\chi,\lambda_{HS})$ and not on the photon-bath temperature $T$. We adopt this approximation because it renders $\langle\sigma v\rangle$ analytic, which is essential to scan the two-dimensional $(m_\chi,\lambda_{HS})$ plane across the full range of UG+NSC parameters  $(\kappa,T_\text{end},x,\omega)$ explored in this work. Closer to the Higgs resonance and the SM thresholds, the velocity expansion of $\langle\sigma v\rangle$ breaks down, and a dedicated treatment is required \cite{Griest:1990kh}, which we leave for future work.
This restriction does not affect the qualitative conclusions of our analysis, since the impact of the energy diffusion on the WIMP parameter space, discussed in Subsection~\ref{subsec:WIMPParameters}, is already captured within the mass range considered here. This allows us to reinterpret every result of Subsection~\ref{subsec:WIMPParameters} directly in the $(m_\chi,\lambda_{HS})$ plane instead of $(m_\chi,\langle\sigma v\rangle)$, which is the standard presentation adopted in the DM direct detection literature, since the same portal coupling $\lambda_{HS}$ that fixes the relic abundance through freeze-out also determines the spin-independent $\chi$-nucleon scattering cross section probed by direct detection experiments such as LZ~\cite{LZ:2022lsv}, against which we compare our results below.

\begin{figure}
\centering
\includegraphics[width=0.48\textwidth]{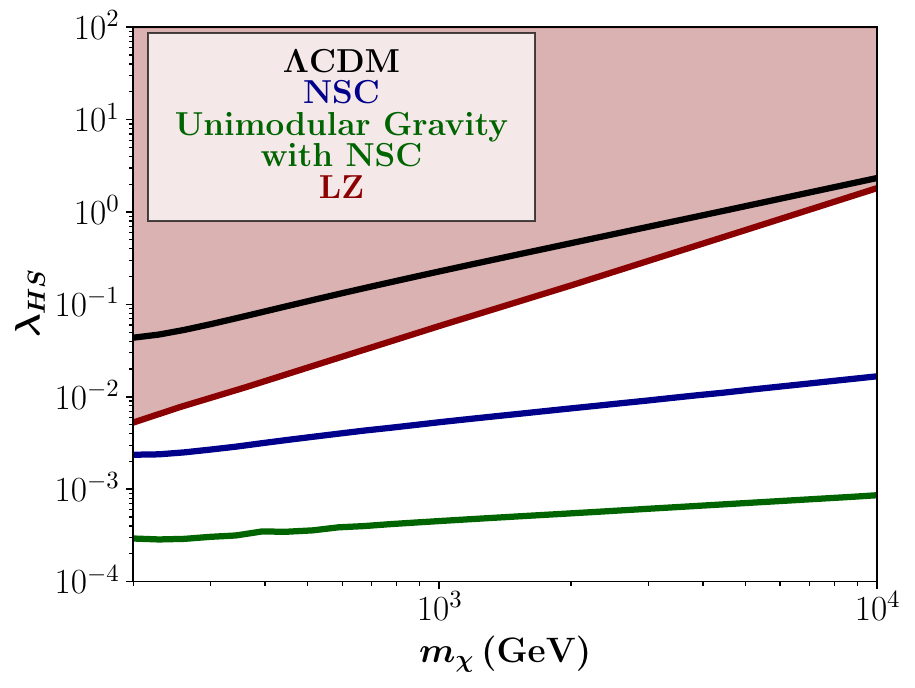}
\caption{$m_\chi$-$\lambda_{HS}$ parameter space for Singlet Scalar DM for $T_\text{end}=7\times10^{-3}$ GeV, $\kappa=10^{-2}$ $\omega=0$, and $x=0.1$. The black, blue, and green lines correspond to the parameters that reproduce the current DM relic density in $\Lambda$CDM, NSC, and UG+NSC scenarios, respectively. The red line is the limit in LZ, and the red region is excluded.}
\label{comparacionscalarLZ}
\end{figure}

Figure~\ref{comparacionscalarLZ} shows the $m_\chi$-$\lambda_{HS}$ parameter space that reproduces the observed DM relic density in the $\Lambda$CDM, NSC, and UG+NSC scenarios (black, blue, and green lines, respectively), together with the current exclusion limit from LZ (red line). As in the generic WIMP case, energy diffusion in UG shifts the relic-density curve toward smaller values of $\lambda_{HS}$ for a given $m_\chi$ relative to $\Lambda$CDM and NSC scenarios, since the additional entropy injected by $\phi$ requires a weaker portal coupling to reproduce the same relic abundance. As a result, a sizable region of the UG+NSC curve remains below the LZ exclusion line in a mass range where the $\Lambda$CDM and NSC curves are already ruled out, indicating that energy diffusion can reconcile a broader range of Singlet Scalar masses and couplings with both the relic density and direct detection constraints.

\begin{figure}
\centering
\includegraphics[width=0.48\textwidth]{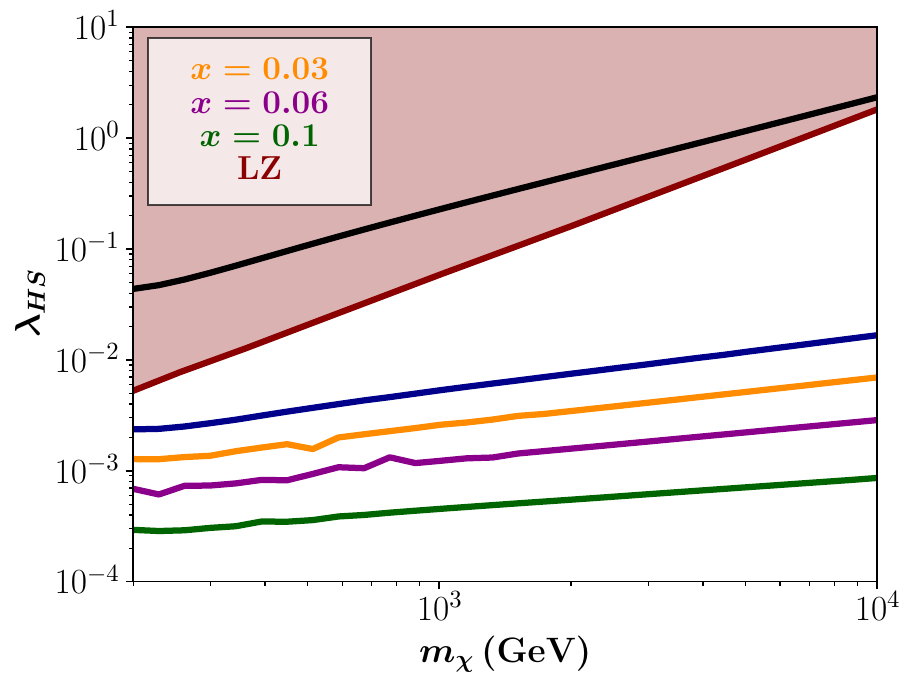}
\caption{$m_\chi$-$\lambda_{HS}$ parameter space for Singlet Scalar DM for $T_\text{end}=7\times10^{-3}$ GeV, $\kappa=10^{-2}$, and $\omega=0$. The black and blue lines correspond to the parameters that reproduce the current DM relic density in $\Lambda$CDM and NSC scenarios; meanwhile, the orange, magenta, and green lines correspond to the space that reproduces the current DM relic density in UG+NSC for $x=0.03$, $x=0.06$, and $x=0.1$, respectively. The red line is the limit in LZ, and the red region is excluded.}
\label{comparacionscalarLZvarx}
\end{figure}

Figure~\ref{comparacionscalarLZvarx} illustrates this effect further by varying the diffusion parameter $x$ at fixed $\kappa$ and $T_\text{end}$. Increasing $x$ progressively lowers the relic-density curve in the $(m_\chi,\lambda_{HS})$ plane, opening additional parameter space below the LZ limit that is inaccessible in the $\Lambda$CDM and NSC scenarios. This shows that even moderate values of the diffusion parameter can appreciably relax the tension between the relic density requirement and current direct detection bounds, and that this effect is not saturated within the range of $x$ considered, suggesting that future direct detection experiments with improved sensitivity will be needed to probe the full parameter space opened by energy diffusion in this benchmark model.

\section{\label{sec:Conclusions}Conclusions}
In this paper, we explored the inclusion of NSC in the UG framework through a new field $\phi$ incorporated in the early Universe. The diffusion parameter $x$ arising from the UG scenario can be associated with $\phi$, which decays before BBN, recovering the $\Lambda$CDM model without diffusion. We compared the parameter space of the model and DM in NSC and UG+NSC, observing that this novel scenario modifies the parameters that reproduce the current DM relic density, as seen in Figs. \ref{ComparacionUGNSCw-25}--\ref{ComparacionNSCUG}.

If DM is detected through any signal, its physical properties (mass and interactions) must be consistent with the cosmological history of $\Lambda$CDM. If this is not the case, new cosmological scenarios must be considered to account for the particle. In this context, varying the $\kappa$, $T_\text{end}$, and $x$ parameters in UG+NSC for fixed DM parameters offers different possibilities to achieve the currently observed DM relic density, as shown in Figs. \ref{varmsvcte}--\ref{ComparacionUGx}. In particular, higher values of $x$ shift the curves downward, allowing lower regions of the model parameter $\kappa$ within the same range of $T_\text{end}$, which translates into lower values of the initial energy density for $\phi$ (i.e., radiation could dominate strongly over $\rho_\phi$ in RI). 

On the other hand, fixing the model parameters $\kappa$--$T_\text{end}$, and varying $m_\chi$, $\langle\sigma v\rangle$, and $x$ opens new windows to explore for WIMP candidates, as seen in Figs. \ref{variacionkappadm}--\ref{variacionomegawimp} reaching parameter values that are excluded in the $\Lambda$CDM scenario. As can be seen, increasing the value of $x$ changes the slope in the $m_\chi$--$\langle\sigma v\rangle$ plane, making it more pronounced and thereby achieving lower values in the thermally averaged cross section for light DM masses. Another interesting behavior is that higher values of $\kappa$ ($\phi$ always dominates over radiation) shift the DM parameters to the left, which means that in UG+NSC, lower values of $\kappa$ achieve higher masses for DM candidates, leaving lighter candidates for scenarios in which $\rho_\phi>\rho_\gamma$, similar to what happens when $T_\text{end}$ decreases, allowing lighter DM candidates. Meanwhile, for different values of $\omega$, a wide range of $\langle\sigma v\rangle$ can be achieved within the same DM mass range, with cases where $\omega\geq0$ being more favored to obtain a heavier DM candidate. These results are entirely general, as they depend only on $m_\chi$ and $\langle\sigma v\rangle$ and are independent of the underlying particle physics realization of the WIMP.

As a concrete application of this general framework, we specialized to the Real Singlet Scalar extension of the SM as a benchmark WIMP realization, restricting our analysis to $m_\chi\gtrsim200$ GeV, i.e., above the top-quark threshold and away from the Higgs resonance and the $W$, $Z$, $h$, and $t\bar{t}$ production thresholds, where $\langle\sigma v\rangle$ can be treated, to excellent approximation, as an analytic function of $(m_\chi,\lambda_{HS})$ alone, independent of temperature. Confronting the resulting $(m_\chi,\lambda_{HS})$ parameter space with the current exclusion limit from the LZ experiment (Fig.~\ref{comparacionscalarLZ}), we found that energy diffusion in UG+NSC opens a sizable region of couplings that remains unconstrained by LZ in a mass range already excluded in the $\Lambda$CDM and some NSC scenarios. Varying the diffusion parameter $x$ (Fig.~\ref{comparacionscalarLZvarx}) further enlarges this allowed region, showing that even a modest amount of energy diffusion can relax the tension between the observed relic density and current direct detection bounds by up to an order of magnitude in $\lambda_{HS}$, consistent with the generic WIMP behavior discussed above.

While our analysis of the Singlet Scalar case was restricted to a single benchmark model and to the high-mass regime where $\langle\sigma v\rangle$ is well approximated as temperature-independent, the qualitative impact of energy diffusion on the relic density and direct detection prospects is expected to hold for other WIMP realizations. Extending this analysis to the resonance and threshold region, as well as to other minimal WIMP models, is left for future work.

\section*{\label{sec:Acknowledgments}Acknowledgments}
C.B. was supported by ANID-Chile under the grant ANID FONDECYT/Regular 1241855. C.B. acknowledges support from the ICTP through the Associates Programme (2023-2028).
C.B. and E.G. acknowledge Vicerrectoría de Investigación y Desarrollo Tecnológico (VRIDT) at Universidad Católica del Norte (UCN) for the scientific support provided by Núcleo de Investigación en Simetrías y la Estructura del Universo (NISEU-UCN), Resolución VRIDT N°200/2025.

\bibliographystyle{apsrev4-2}
\bibliography{bibliography}

\end{document}